\documentclass[twocolumn]{aastex61}
\usepackage{natbib}
\usepackage{subfigure}
\usepackage{amsfonts}
\usepackage{tabularx}
\usepackage{tcolorbox}
\usepackage{color}
\usepackage{graphicx}
\usepackage{amsmath}
\usepackage{amssymb}
\usepackage{mathrsfs}

\newcommand{\mat}[1]{\boldsymbol{\mathbf{#1}}}

\newcommand{\m}{\ensuremath{\,{\rm m}}}

\newcommand{\Mpc}{\ensuremath{\,{\rm Mpc}}}
\newcommand{\K}{\ensuremath{\, {\rm K}}}
\newcommand{\mK}{\ensuremath{\, {\rm mK}}}

\newcommand{\MHz}{\ensuremath{\, {\rm MHz}}}

\begin{document}

\title{Extracting 21cm signal by frequency and angular filtering}

\author{Qizhi Huang}
\affil{Key Laboratory of Computational Astrophysics, National Astronomical Observatories, Chinese Academy of 
Sciences, Beijing 100101, China}
\affil{School of Astronomy and Space Science, University of Chinese Academy of Sciences, Beijing 100049, China}
\affil{Universit\'e Paris-Sud, LAL, UMR 8607, F-91898 Orsay Cedex, France $\&$ CNRS/IN2P3, F-91405 Orsay, France}

\author{Fengquan Wu}
\affil{Key Laboratory of Computational Astrophysics, National Astronomical Observatories, Chinese Academy of 
Sciences, Beijing 100101, China}

\author{Reza Ansari}
\affil{Universit\'e Paris-Sud, LAL, UMR 8607, F-91898 Orsay Cedex, France $\&$ CNRS/IN2P3, F-91405 Orsay, 
France}

\author{Xuelei Chen}
\affil{Key Laboratory of Computational Astrophysics, National Astronomical Observatories, Chinese Academy of 
Sciences, Beijing 100101, China}
\affil{School of Astronomy and Space Science, University of Chinese Academy of Sciences, Beijing 100049, China}
\affil{Center for High Energy Physics, Peking University, Beijing 100871, China}

\shortauthors{Huang et al.}


\begin{abstract}
Extracting the neutral hydrogen (HI)  signal is a great challenge for cosmological
21cm experiments, both the astrophysical foregrounds and the receiver noise are typically
several orders of magnitude greater than the 21cm signal. However, the different properties of the 21cm signal, 
foreground, and noise can be exploited to separate these components. The foregrounds are generally smooth or
correlated over the frequency space along a line of sight (l.o.s.), while both the 21cm signal and the noise varies
stochastically along the same l.o.s. The foreground can be removed by filtering out the smooth component in the frequency
space.  The receiver noise is basically uncorrelated for observations at different times, hence for surveys
they are also uncorrelated in the different directions, 
while the 21cm signal which traces the large scale structure are correlated up to certain scales.
In this exercise, we apply Wiener filters in frequency and angular space, to extract the 
21cm signals. We found that the method works well. The inaccurate knowledge about beam 
could degrade the reconstruction, but the overall result is still good, showing that the method is fairly robust.

\end{abstract}

\keywords{}


\section{Introduction}

The neutral hydrogen 21cm line of  is one of the most promising tools to study the observable universe. 
Tomographic observation of the redshifted 21cm line could be used to reveal the evolution of the 
intergalactic medium (IGM) throughout the Epoch of Reionization (EoR) \citep{1997ApJ...475..429M,2003ApJ...596....1C}, 
to map out the large scale structure and constrain the cosmological parameters  including the dark energy
equation of state \citep{2008PhRvL.100i1303C,2008PhRvD..78b3529M}. In principle, 
it could even probe the cosmic dark age \citep{Loeb:2003ya}.
Comparing with the cosmic microwave background (CMB), which images the 
Universe at the last scattering surface during the epoch of recombination, 
one advantage of the redshifted 21cm tomography signal as a cosmological probe
is that it provides three dimensional (3D) map of the Universe at different redshifts, giving 
more information and also showing how the Universe evolved.
	
In recent years, a number of experiments set the 21cm observation as one of their main scientific goals, 
for example the experiment with existing telescopes such as the
GBT (the Green Bank Telescope; \citealt{2010Natur.466..463C, 2013ApJ...763L..20M, 2013MNRAS.434L..46S}), 
GMRT (the Giant Metrewave Radio Telescope; \citealt{2011MNRAS.413.1174P}),  and telescopes newly 
built or being built, such as the 21CMA \citep{2016ApJ...832..190Z}, Tianlai (\citealt{2012IJMPS..12..256C}), 
BINGO (BAO from Integrated Neutral Gas Observations; \citealt{2013MNRAS.434.1239B}), 
LOFAR (the LOw-Frequency Array; \citealt{2013A&A...556A...2V}), 
MWA (the Murchison Widefield Array; \citealt{2013PASA...30....7T}), 
PAPER (the Precision Array for Probing the Epoch of Re-ionization; \citealt{2010AJ....139.1468P}), 
CHIME (the Canadian Hydrogen Intensity Mapping Experiment; \citealt{2014SPIE.9145E..22B}), 
HERA (the Hydrogen Epoch of Reionization Array; \citealt{2017PASP..129d5001D}), 
FAST \citep{2011IJMPD..20..989N} and the SKA \citep{2013arXiv1311.4288H}, etc. 
	
Although the redshifted 21cm line can provide a large amounts of information for cosmology, its detection is difficult. 
The cosmological 21cm signal whose brightness temperature $\sim$0.14 mK at redshift $z$$\sim$0.8 
\citep{2010Natur.466..463C, 2013ApJ...763L..20M} is highly contaminated by  the 
foreground emissions whose brightness temperatures are 4--5 orders of magnitude higher 
\citep{2008MNRAS.388..247D,2012MNRAS.419.3491L}. 
The foregrounds at low frequency include primarily the Galactic synchrotron emission which is originated from the cosmic ray
electrons moving in the Galactic magnetic field, 
the Galactic free-free emission which is produced by free electrons scattering off ions without being captured, 
extragalactic radio sources such 
as radio-loud galaxies and quasars \citep{1999A&A...345..380S,2008MNRAS.389.1319J}. Additionally,  the observation 
are also affected by the radio frequency interference (RFI) and propagation effects in the ionosphere. 

Removing the foregrounds has been a big challenge in the redshifted 21cm experiments, and a number of 
methods have been proposed and developed. The general idea is to exploit the different properties of foregrounds 
and cosmological 21cm signal. Along one line of sight (LoS), the cosmological 21cm signal varies with redshift or frequency 
randomly, while the the astrophysical foregrounds vary smoothly.
The foreground can in principle be removed by light-of-sight fitting \citep{2006ApJ...650..529W,2008MNRAS.391..383G} 
or by cross-correlating 
different frequency bins data \citep{2005ApJ...625..575S}. More recently, blind or semi-blind methods such as 
the Singular Value Decomposition (SVD) method \citep{2011MNRAS.413.1174P,2013ApJ...763L..20M}, 
Robust Principle Component Analysis \citep{Zuo:2018gzm}, and 
Independent Component Analysis \citep{2012MNRAS.423.2518C,2014MNRAS.441.3271W},  are applied.

In this paper, we introduce a simple and fast method which is based on the Wiener filter to extract the 21cm signal. 
The Wiener filter has been widely used in signal processing,  especially for removing noise in time series or
images. We use a simulation to demonstrate our method.

The rest of this paper is structured as follows. In Sec. 2, we describe the Wiener filter method in general, and also describe
the set up  of our simulation. In Sec. 3, we apply the method to the simulated data, and present the results we obtained. 
Finally, we discuss a number of relevant issues in the data processing and concludes in Sec. 4.


\section{Method}
\label{sec:method}
 
In a HI intensity mapping survey experiment, the  observed sky emission is a mixture of the 21cm signal, 
noise and foregrounds, to extract the cosmologically interesting 21cm signal, 
their different statistical properties are used to separate them.
On the relevant scale, the cosmological 21cm signal varies stochastically, as the signal strength at each 
frequency corresponds to the emission of a specific redshift, and on large scales the density at each different
position are independent to each other (though there is correlation to some degree).  
By comparison, the foreground varies smoothly in frequency space, so it could in principle be distinguished from the 
21cm line. In addition to the sky signals, the electronic circuit of the receiver also generate noise. 
After bandpass calibration, the noise could be approximated as zero-mean white noise. 
Based on their different properties, it is in possible, at least in principle, to extract the 
21cm signal from the much stronger foregrounds and noise. 

In an actual radio telescope, the data would be first pre-processed to remove bad data (e.g. those with 
hardware malfunction or radio frequency interference), calibrated, re-binned in frequency and time resolution, 
re-arranged in predefined order, then used to form images cubes with 
two angular dimension and one frequency dimension. These steps will depend on the particular telescope 
in question, e.g. the data of a single dish telescope would be processed very differently from the data the of 
an interferometer array. Here, we will deal the data processing after these steps. We shall assume that an 
image cube have been obtained through these steps.

Below we make the 21cm signal extraction in two steps. In the first step, we filter the data along each l.o.s, 
to remove the foreground component by using their different properties. As a result, the foreground data
could be significantly suppressed, while the 21cm signal and the thermal noise is kept.
In the second step, we filtering the data in the 2D angular space, to remove the randomly fluctuating 
noise signal, while keeping the more stable and consistent 21cm signal. 

%
%
 
\subsection{Wiener Fileter}
We assume that in an experiment  the observational data $\mat{y}$ is linearly related to $\mat{x}$, the 
physical quantity we try to measure, 
\begin{equation}
	\mat{y} = \mat{A}\, \mat{x} + \mat{n}
\label{eq:y}
\end{equation}
where $\mat{A}$ is the response matrix of the system, and $\mat{n}$ is random noise. We use boldface letter
to denote vectors and matrices, and  $^T$ denotes matrix or vector transpose (we assume the data are real numbers here).
The covariance matrices for the signal and noise are
 $\mat{S}=\langle \mat{x}\, \mat{x}^T \rangle$, $\mat{N}=\langle \mat{n}\, \mat{n}^T \rangle$ respectively.
If $\mat{S}$ and $\mat{N}$ are known, an unbiased estimate of the signal can be obtained by applying a 
 Wiener filter $\mat{\mat{W}}$ to the  data \citep{Tegmark:1996qs}, 
\begin{equation}
\hat{\mat{x}} = \mat{W} \mat{y} \equiv \mat{S} \mat{A}^{T} \left[\mat{A} \mat{S} \mat{A}^T + N \right]^{-1} \mat{y}
\label{eq:wiener}
\end{equation}
The Wiener filter is optimal in the sense that it minimizes the variance of the estimator 
$$V=\langle (\mat{x}-\hat{\mat{x}})(\mat{x}^{T}-\mat{\hat{x}^T}) \rangle$$.

\subsection{Simulation Setup}
As the signal extraction utilizes our prior knowledge or expectation of the 21cm signal, foreground and noise, 
the optimal filter depends on the statistics of them, so the property of the filter also depends on the particular problem. 
Here we describe the basic setup we use in this exercise.

We consider the the mid-redshift experiment aimed at detecting the baryon acoustic oscillation (BAO) signal of 
the large scale structure, such as the ongoing survey projects on 
GBT \citep{2013ApJ...763L..20M,2013MNRAS.434L..46S}, and dedicated experiments such as 
Tianlai \citep{2012IJMPS..12..256C,Xu:2014bya,Zhang:2016whm,Zhang:2016miz}, 
CHIME \citep{2014SPIE.9145E..22B}, BINGO \citep{2014arXiv1405.7936D} and HIRAX \citep{2016arXiv160702059N}.
For such projects, the (synthetic) aperture is $50 \sim100$ m. Below, we shall consider a fiducial frequency of 
800 MHz ($z=0.7755$) and 100 m aperture, the corresponding full width half maximum (FWHM) 
resolution of $0.25^\circ$.

We generate the 21cm signal as follow. We adopt the Planck 2015 \citep{2016A&A...594A..13P} 
best fit cosmological parameters for our fiducial model. For  $z= 0.7755$ 
the corresponding comoving angular diameter distance is 
$D_A = 1068.95 \Mpc~h^{-1}$. We consider a frequency resolution of 0.1 MHz, the corresponding to 
a comoving radial distance interval of $\Delta D_c=0.43 \Mpc ~h^{-1}$ at this redshift. 
For our 21cm simulation, we generate an image cube with $200^3$ voxels, which is convenient for
computation. Then the frequency interval is 20 MHz, i.e. $790 \sim 810 \MHz$. The corresponding 
angular size per pixel is $\Delta \theta = \Delta D_c / D_A = 0.023^{\circ}$, smaller than the beam FWHM, 
which is good as our computation precision would not be affected by too large pixel sizes. 
The whole box has an angular area of  
 $4.6^{\circ} \times 4.6^{\circ}$.  The volume of the corresponding 3D box in real space is 
$V = L^3 = (86\,{\rm Mpc}\,h^{-1})^3$.

We assume a thermal noise with $\sigma_n=200$ mK per voxel, 
which is $\sim 10^3$ times of the 21cm fluctuation $\sim$0.2 mK in our frequency (redshift) range. 
Note that the beam area is about $10^2$ times of the pixel area, so if the pixel noise is independent of each other, 
this corresponds to a noise level of  $\sigma_T =20 \mK$ per beam. Note that using the measurement equation
\begin{equation}
\sigma_T \sim \frac{T_{\rm sys}}{\eta \sqrt{\Delta \nu t}},
\label{eq:meaEq}
\end{equation}
where $\eta < 1$ is the efficiency of the system, and take $\Delta \nu =0.1\MHz$.
If the system is stable with a typical system temperature of $T_{\rm sys} =20\K \sim 100 \K$, and the noise is thermal, 
this level of noise can be achieved with a few minutes integration time. However,  in the real telescopes there might be 
a noise floor of non-thermal origin preventing the noise reaching the low thermal value given by Eq.~\ref{eq:meaEq}
no matter how long the integration time, so the actual noise might be higher.

\begin{figure}[htbp]
\centering
	\includegraphics[width=0.46\textwidth]{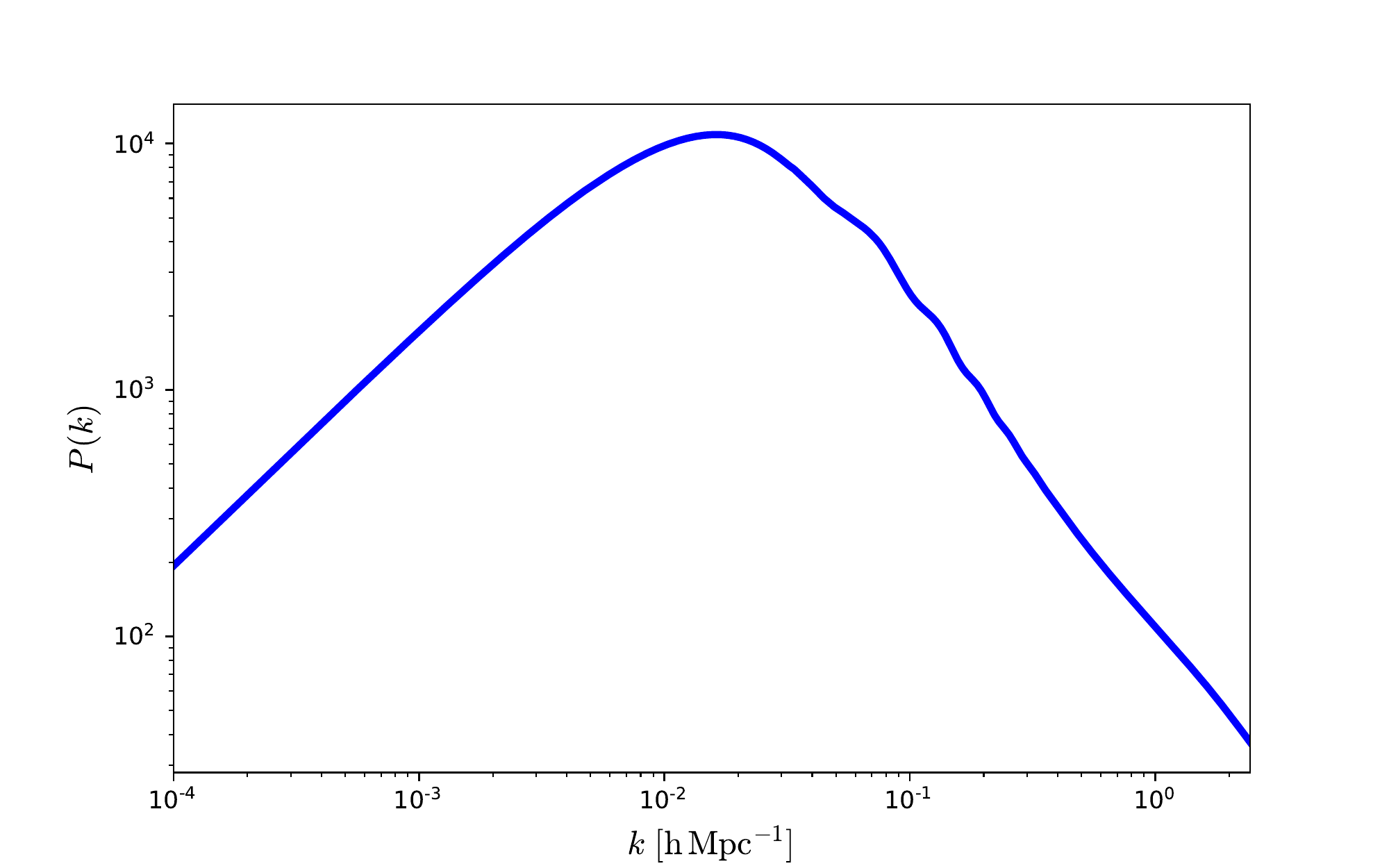}
	\caption{ Power spectrum of dark matter at $z=0.7755$ (with non-linear correction).}
	\label{fig:Pk}
\end{figure}

\begin{figure}[htbp]
\centering
\includegraphics[width=0.46\textwidth]{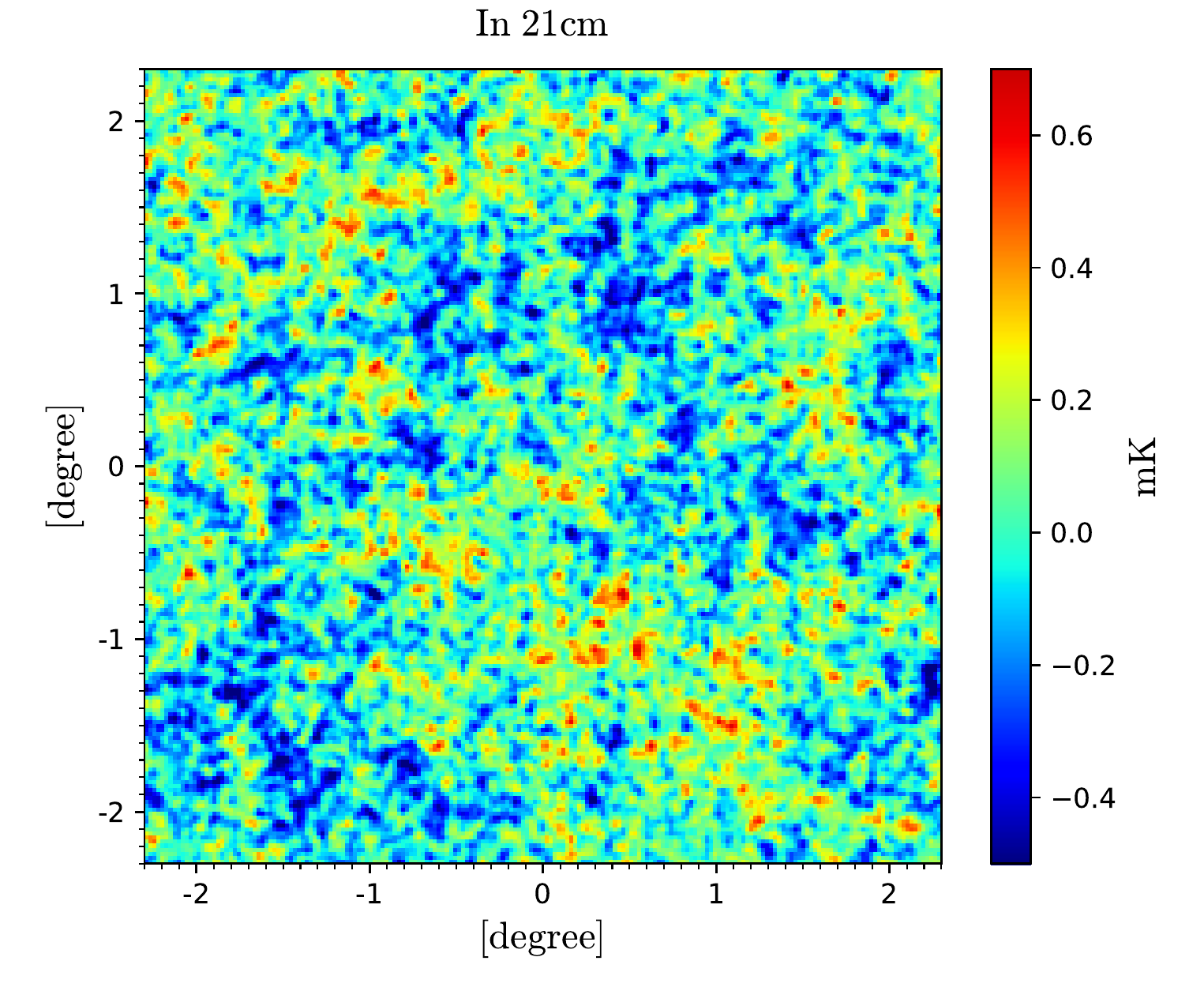}\\
\includegraphics[width=0.46\textwidth]{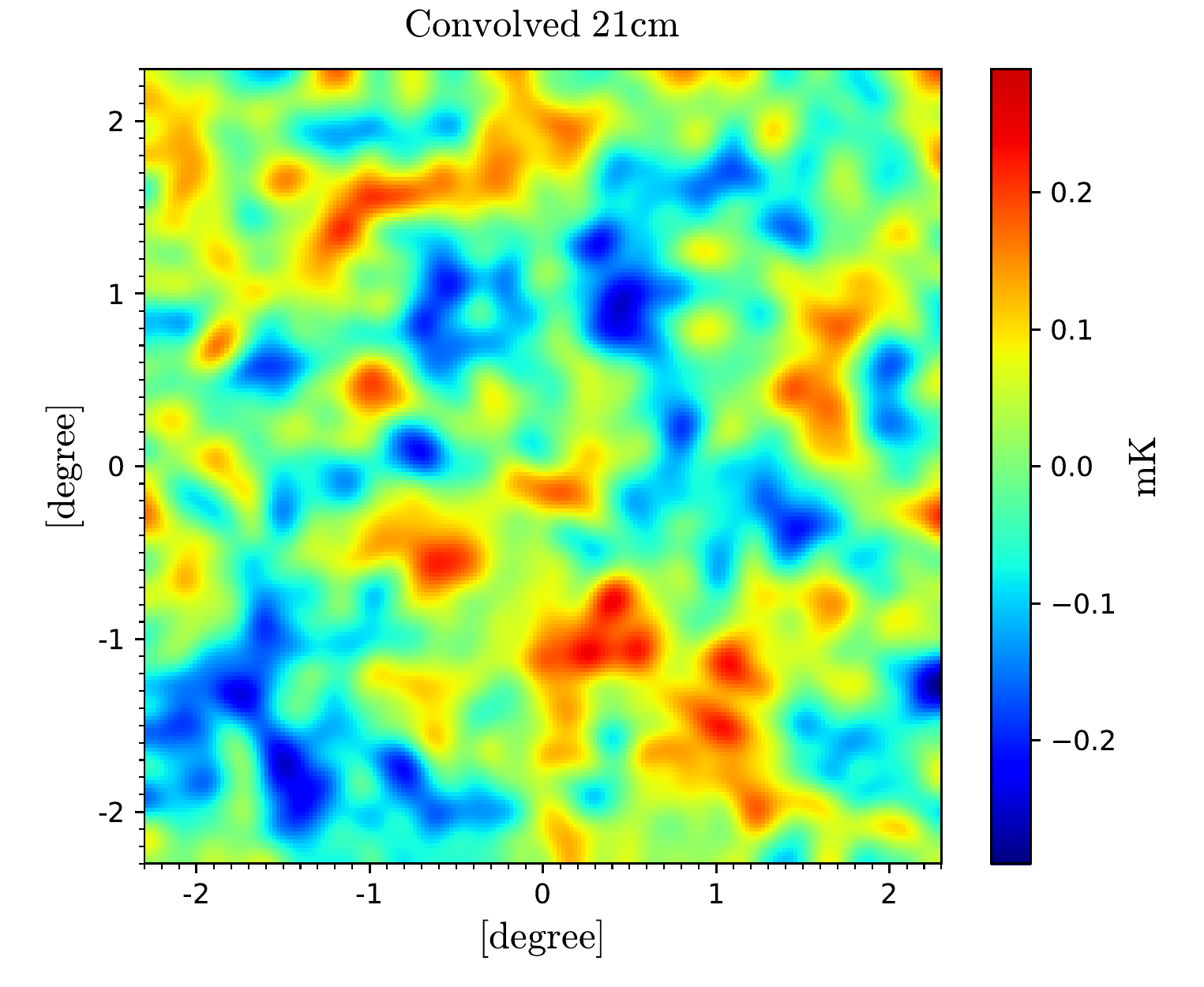}
\caption{Top: the simulated 21cm signal at 800 MHz; bottom: the 21cm signal 
convolved with a FWHM $0.25^{\circ}$ Guassian beam.}
\label{fig:inmap}
\end{figure}

Assuming the neutral hydrogen evolution is linear on the relevant scales, the dark matter power spectrum is shown in 
Fig.~\ref{fig:Pk}.  We then generate random density distribution in the simulation box according to the power spectrum.
Adopting an HI bias of $b_{\rm HI}=0.70$ and HI density ratio 
$\Omega_{\rm HI}=6.6\times10^{-4}$ \cite{2010Natur.466..463C,2013ApJ...763L..20M} , 
the average brightness temperature of 21cm signal around $z$$\sim$0.8 is given by 
\begin{eqnarray}
\bar{T}_{21} &\approx & 0.284 \left(\frac{\Omega_{\rm HI}}{10^{-3}}\right) \left(\frac{h}{0.73}\right) \left(\frac{1+z}{1.8}\right)^{1/2} \nonumber \\
		&& \times \left(\frac{\Omega_m+\Omega_{\Lambda}(1+z)^{-3}}{0.37}\right)^{-1/2}  \mK
\label{eq:T21mean}
\end{eqnarray}
The simulated 21cm fluctuation temperature map with
$\delta T_{21}(\vec{x}) = b_{\rm HI}\bar{T}_{21}  \delta(\vec{x}) $
is shown in the top panel of Fig.~\ref{fig:inmap}. 
We also produce a map convolved with a Gaussian beam whose FWHM resolution is $0.25^{\circ}$, 
corresponding to the resolution of a telescope with an aperture of $\sim 100 \m$ at 800 MHz frequency, this  is
shown in the bottom panel of Fig.~\ref{fig:inmap}. 

We also include a few simple foreground models in the simulation.  At the low frequencies, 
we consider three kinds of spectrum indices of diffuse foreground: the Galactic 
synchrotron emission which dominates the low-frequency sky; a slightly more sophisticated model 
with frequency-varying index; multiple indices which are likely to be produced by additional components such as  
synchrotron emission, free-free emission, etc.
In the simplest case, the brightness temperature of the foreground component is modeled with a single spectral index, 
\begin{eqnarray}
f(\nu) = A_* \left( \frac{\nu}{\nu_*} \right)^{\beta_*}
\label{eq:1index}
\end{eqnarray}
We adopt $A_*=5.3$ K, the average temperature of foreground at 800 MHz \citep{2017MNRAS.464.3486Z}, 
$\beta_*=-2.76$ \citep{1988A&AS...74....7R, 2001A&A...368.1123T}. A slight improvement is to consider 
a running index foreground  given by \cite{2006ApJ...650..529W}
\begin{eqnarray}
f(\nu) = A_* \left( \frac{\nu}{\nu_*} \right)^{\beta_* -0.1\ln(\nu/\nu_*)}
\label{eq:runningindex}
\end{eqnarray}
Finally,  for multiple indices foreground
\begin{eqnarray}
f(\nu) = 5.3 \left( \frac{\nu}{\nu_*} \right)^{-2.76} + 0.2 \left( \frac{\nu}{\nu_*} \right)^{-2.1} + 0.1 \left( \frac{\nu}{\nu_*} \right)^{-3.2}
\label{eq:multiindex}
\end{eqnarray}

\begin{figure}[htbp]
\centering
\includegraphics[width=0.44\textwidth]{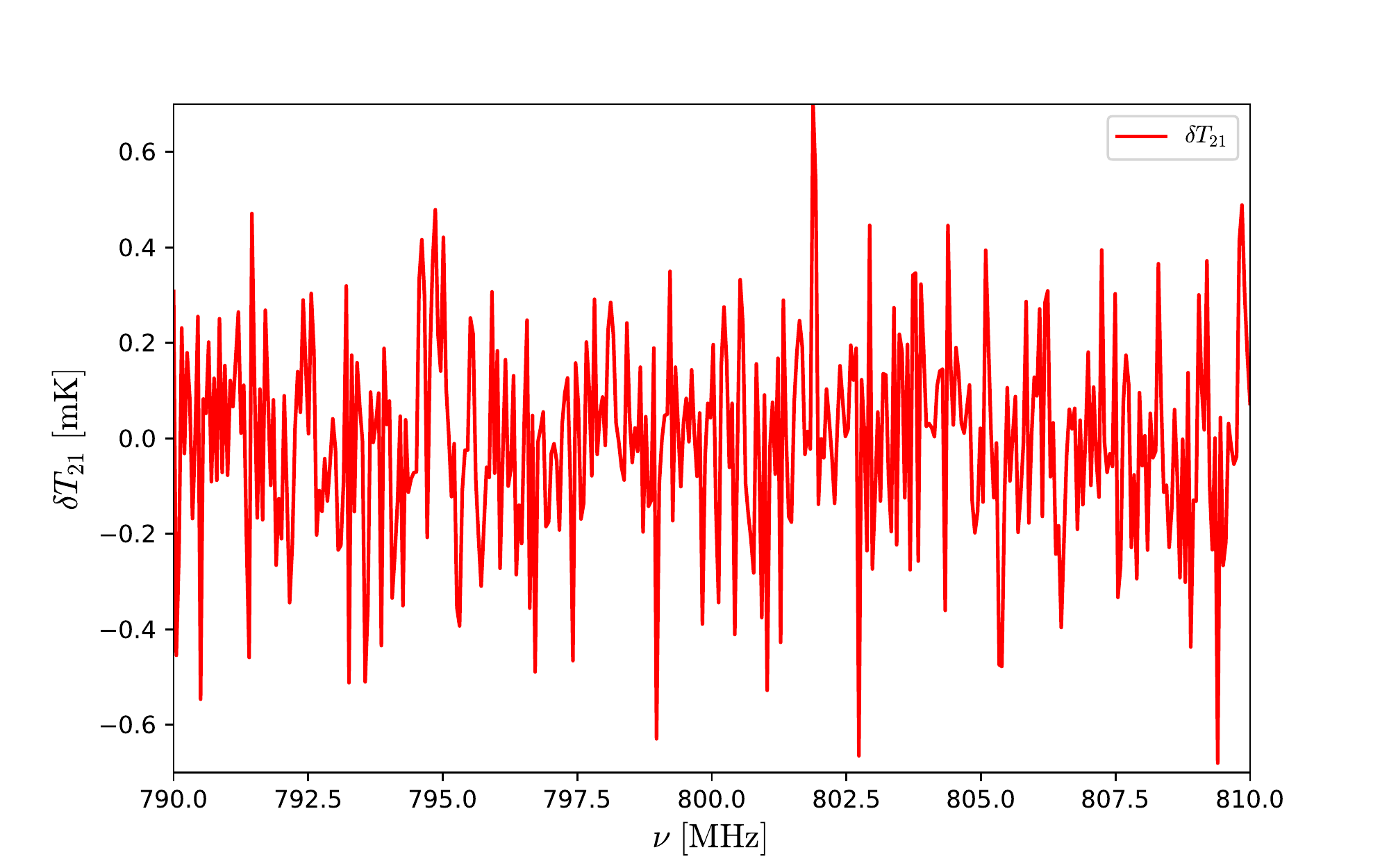}
\includegraphics[width=0.44\textwidth]{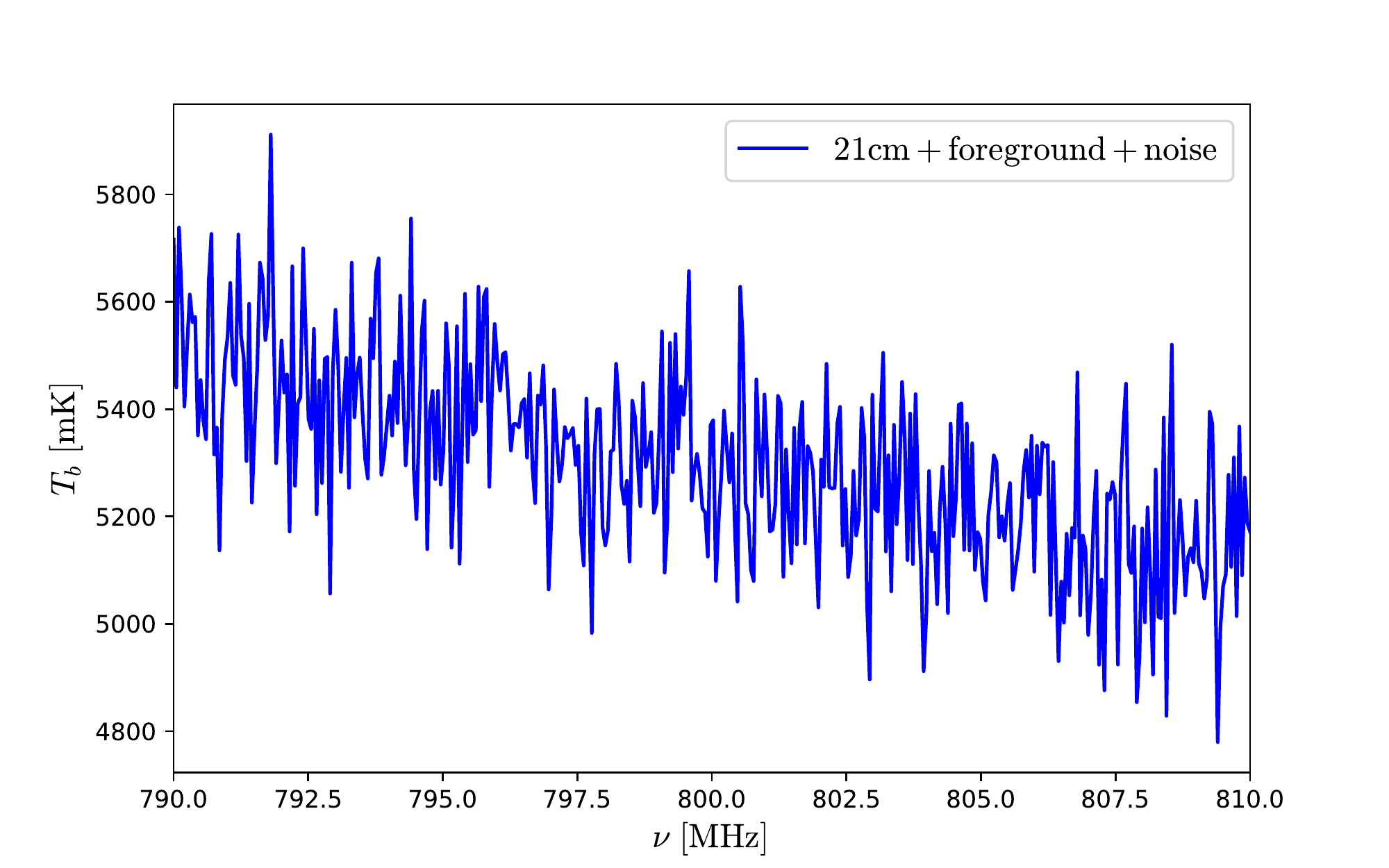}
\caption{ The input spectrum along one line of sight.  
 Top: the input 21cm signal;  Bottom: 21cm signal + single index foreground + noise. }
\label{fig:freq_1index_input}
\end{figure}

In Fig.~\ref{fig:freq_1index_input}, we show the input 21cm signal along one line of sight in the top panel, 
and the the total temperature including the 21cm signal,  the foreground (a single power index component) and randomly
generated noise in the bottom panel.

\section{Results}
Here we apply this method to the extraction of 21cm signal from the observational data with foreground and noise.
In the present paper, we shall adopt a two step procedure: in the first step, we process the data cube along each line of 
sight by removing the smoothly distributed foreground component. In the second step, we apply the Wiener filter in 
the two dimension angular space, which reduce the noise significantly to recover the 21cm signal. 

\subsection{Frequency Filtering}
\label{sec:freqspace}

In frequency space, for the relatively low resolution (0.1 MHz) required for intensity mapping, we may neglect 
the small side lobes in the frequency channels, and take $\mat{A}=\mat{I}$, where $\mat{I}$ is the identity matrix. 
We rewrite Eq.~(\ref{eq:y}) as
	\begin{eqnarray}
		y(\nu) = f(\nu) + s(\nu) + n(\nu)
	\label{eq:ynu}
	\end{eqnarray}
where $f(\nu)$ is the foreground, $s(\nu)$ is the 21cm signal, $n(\nu)$ is the noise, which we assume to be a 
white noise with zero mean. We assume the signal, foreground and noises are uncorrelated with each other, so that 
$\langle \mathbf{(f+s+n)\,(f+s+n)}^T \rangle  =\mat{F} + \mat{S} + \mat{N}$, where $\mat{S}=\langle \mat{s s^T} \rangle $, $
\mat{F}=\langle \mat{f f^T} \rangle $,
 and $\mat{N}=\langle \mat{n n^T} \rangle $.

Along one line of sight,  both the 21cm signal and the noise are stochastically varying on the relevant scales, 
while the foreground is more smooth, so here we may use this to extract the foreground from the data first. 
If we ignore the slight imperfection 
in frequency channels which are fairly small, so that the response can be treated as $\delta$ function, the 
foreground extraction filter is given by $\mat{W}_\nu^f=\mat{F}[\mat{F}+\mat{S}+\mat{N}]^{-1}$, 
while the signal+noise is given by $\mat{W}_\nu = \mat{I}-\mat{W}_\nu^f$.

We note that in the real world the foreground is unknown, so strictly speaking the Wiener filter method can not be applied.
Nevertheless the foreground is believed to be smooth in the frequency space, so that even though the Wiener filter 
constructed this way is not very precise, it could still serve as a low-pass filter to extract the smooth component of the 
data. In fact, we also tried applying a simple low-pass filter and found the result is practically the same.

\begin{figure}[htbp]
\centering
\includegraphics[width=0.23\textwidth]{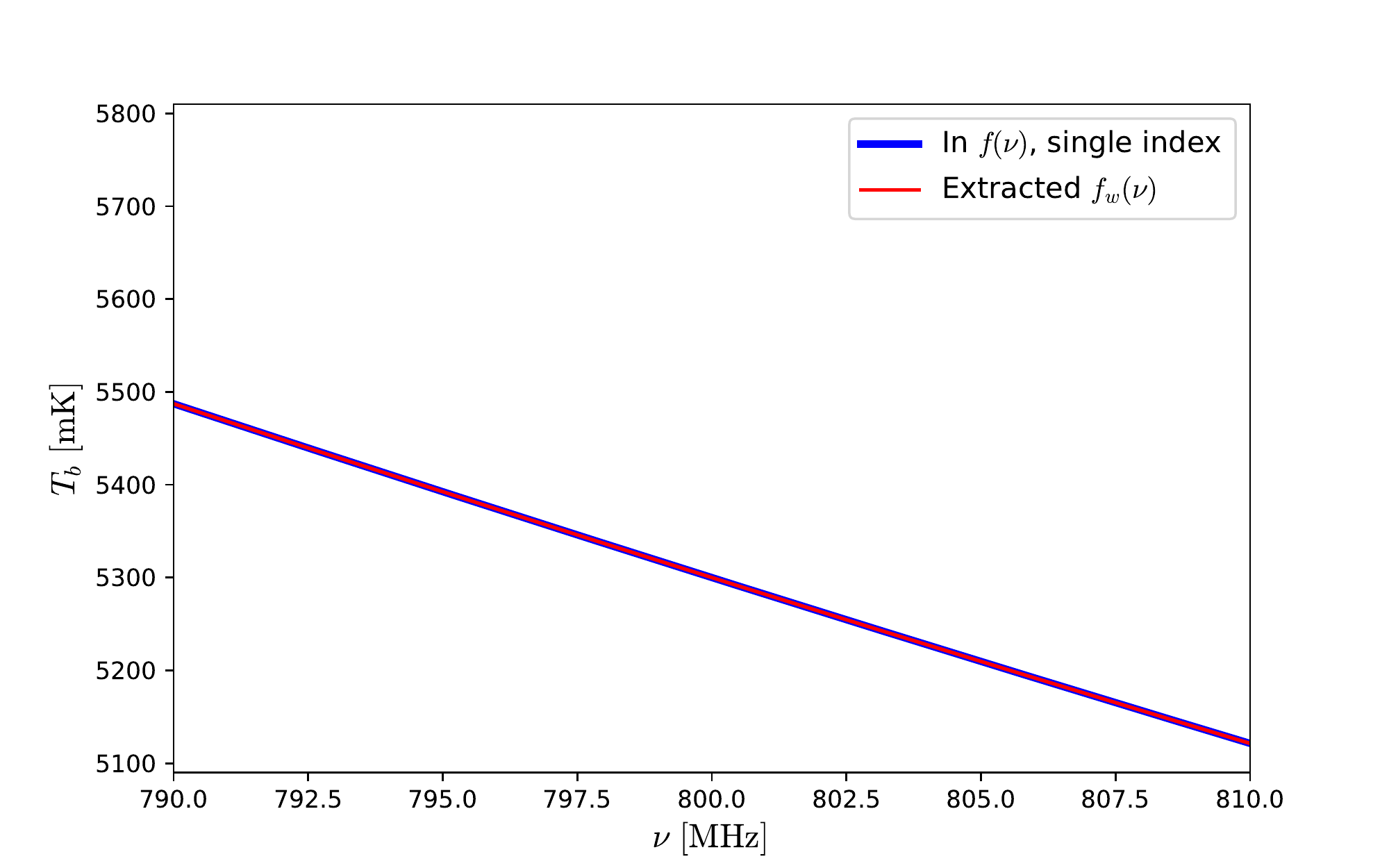}
\includegraphics[width=0.23\textwidth]{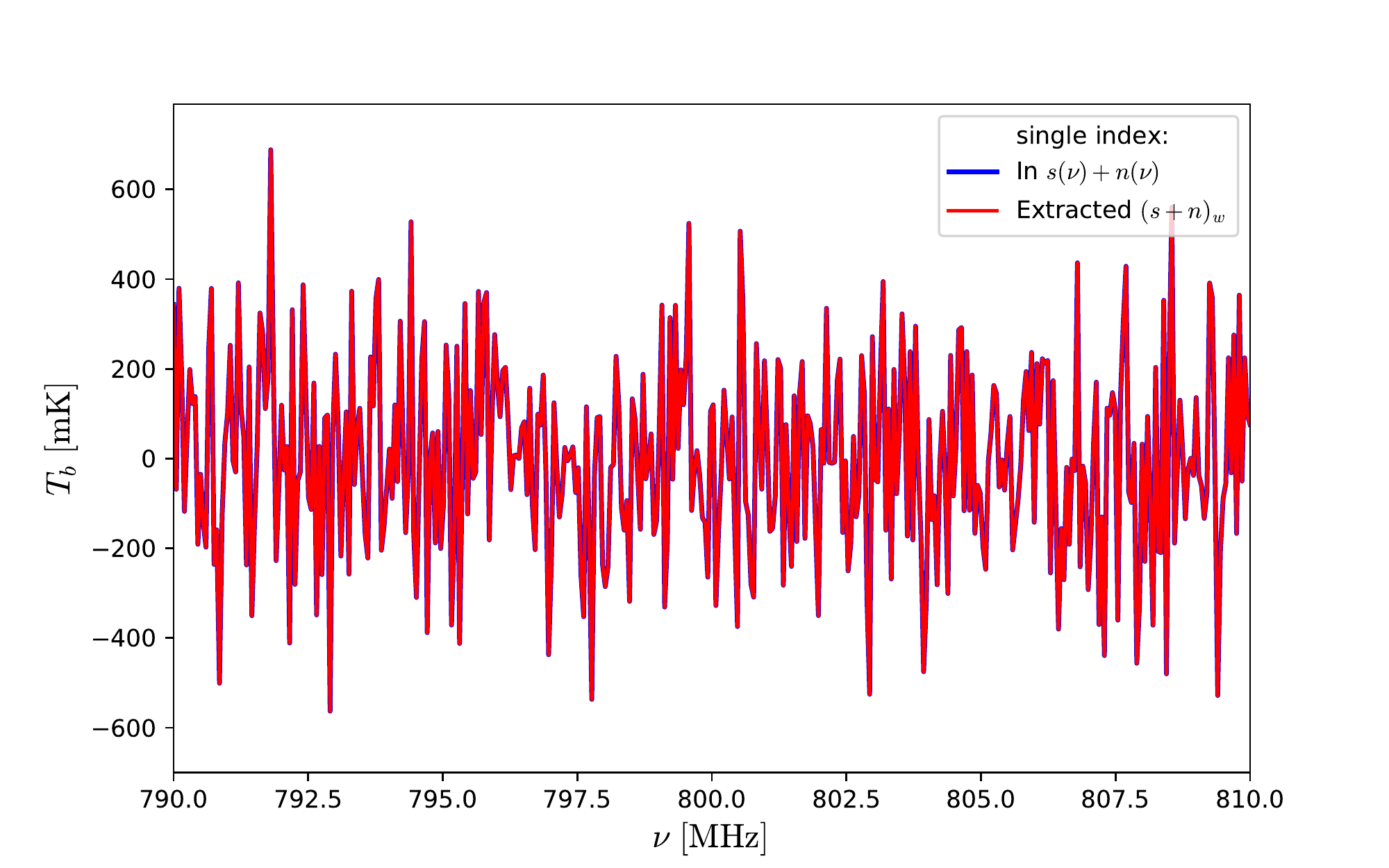} \\
\includegraphics[width=0.23\textwidth]{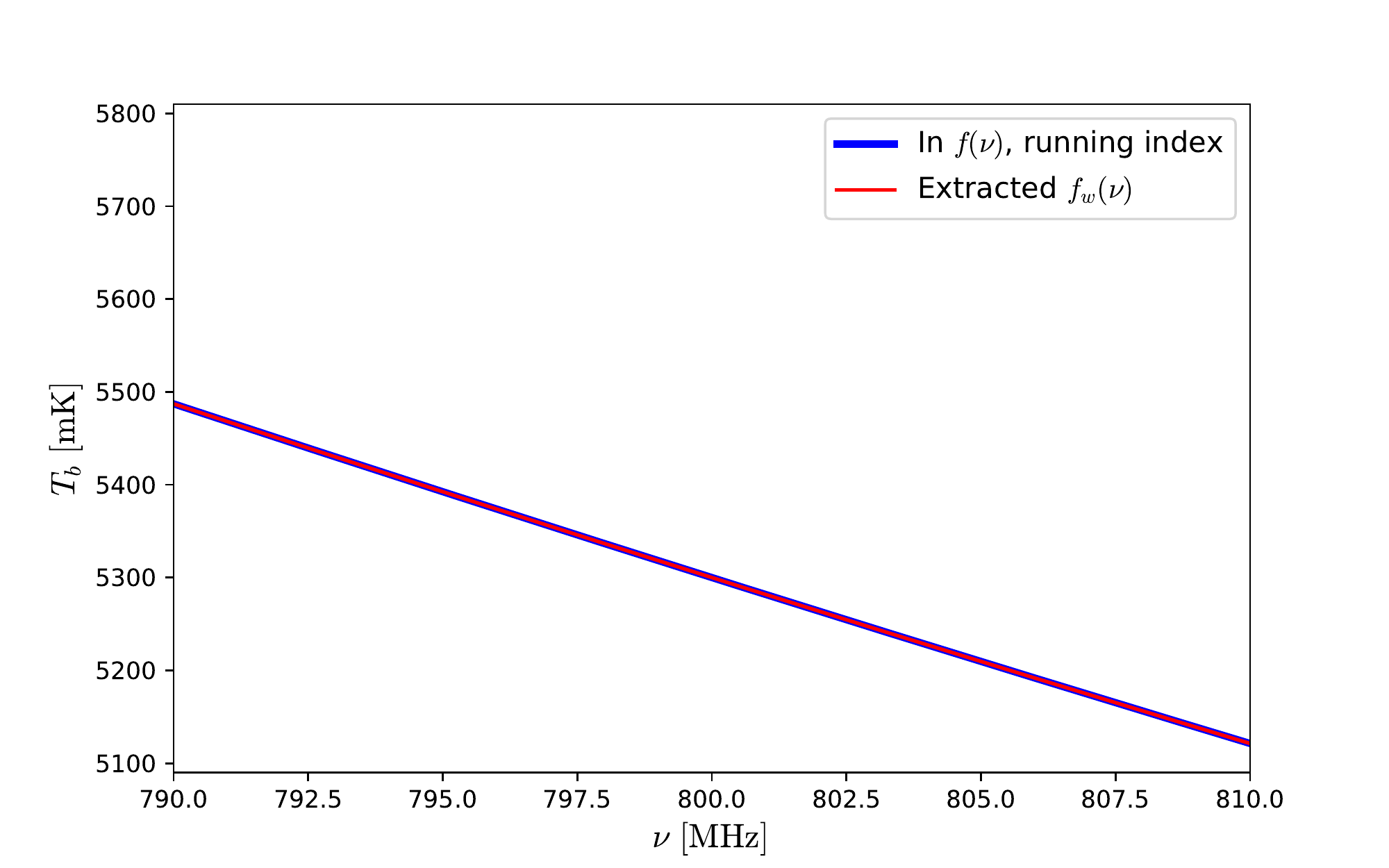}
\includegraphics[width=0.23\textwidth]{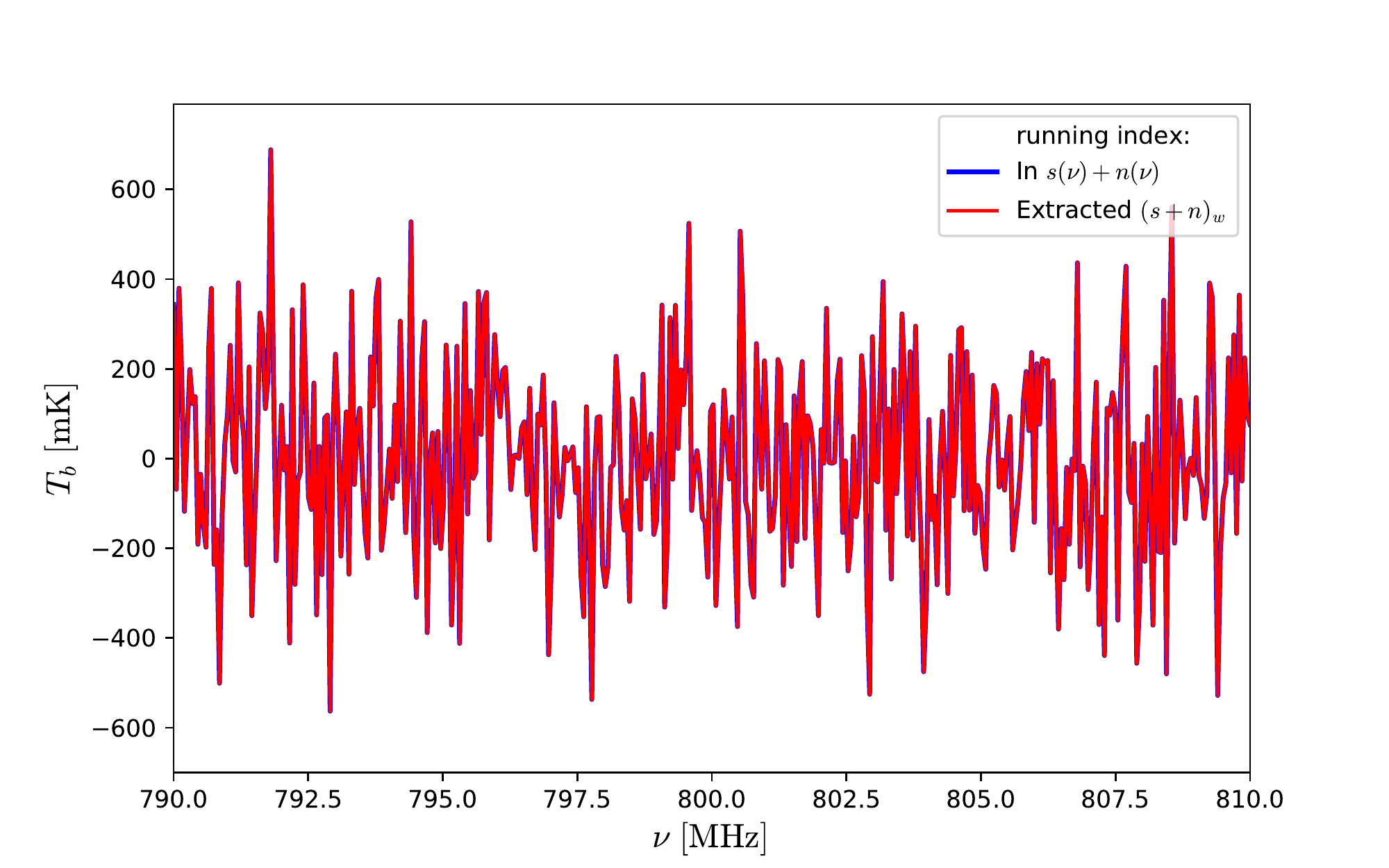} \\
\includegraphics[width=0.23\textwidth]{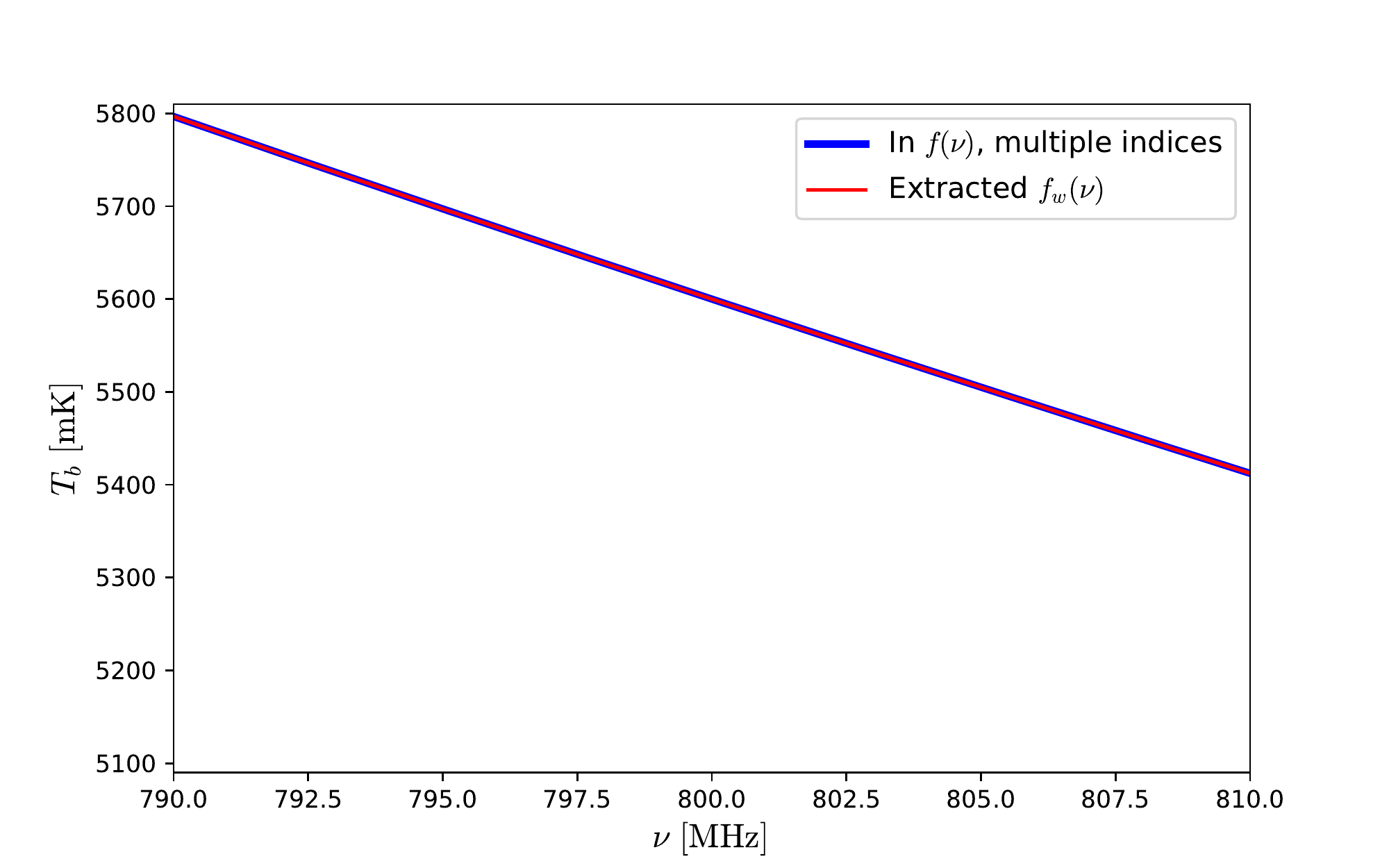}
\includegraphics[width=0.23\textwidth]{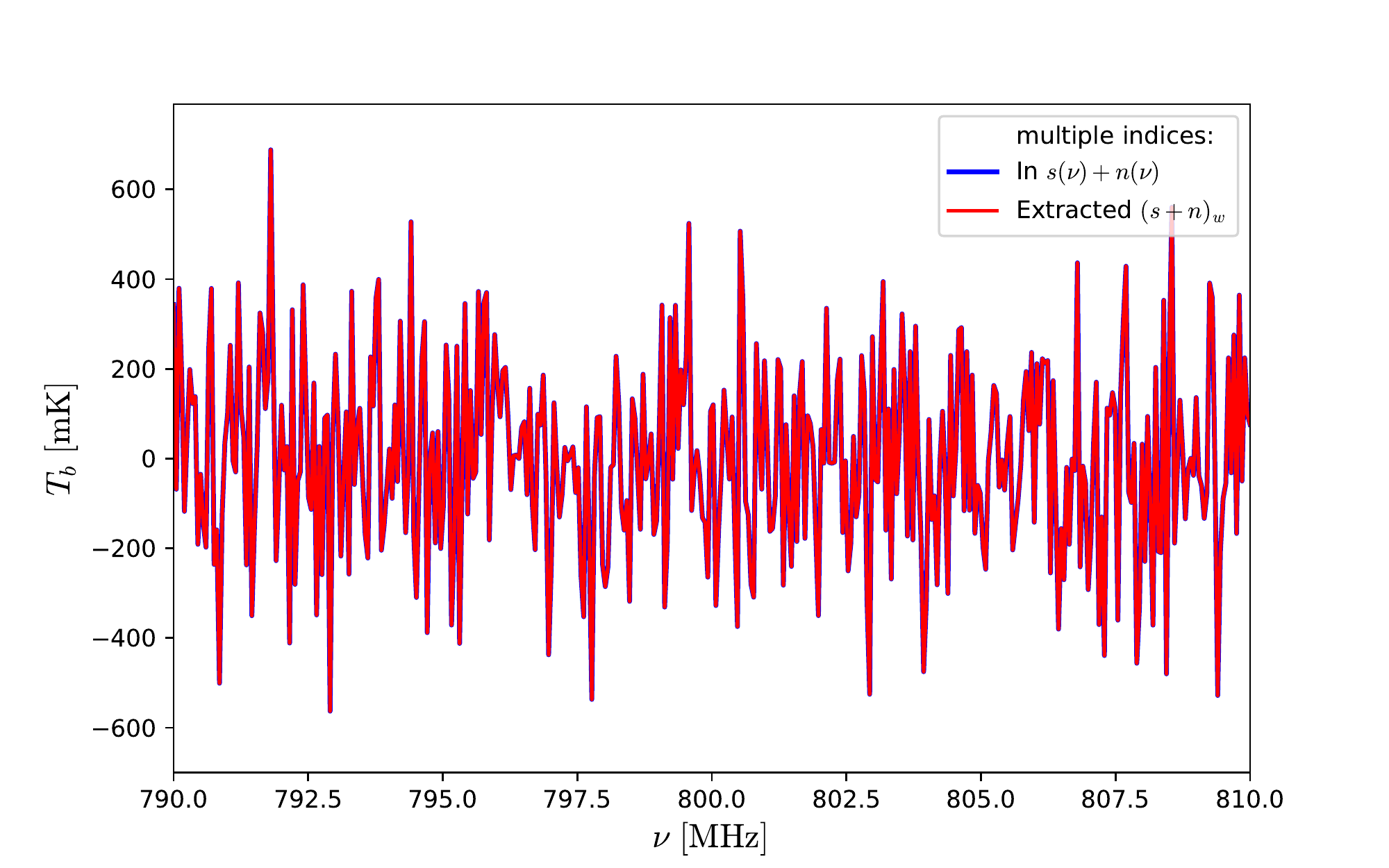}
\caption{ Left: extracted foregrounds ($\mat{W}^f\,\mat{y}$). Right: extracted 21cm+noise ($[\mat{I}-\mat{W}^f]\,\mat{y}$). 
Top to bottom: different foreground models with single index, running index and multiple indices respectively. }
\label{fig:freq_3index}
\end{figure}

The filtered data $\mat{W}^f  \mat{y}$ and $[\mat{I}-\mat{W}^f]\,\mat{y}$ are shown for a random line of sight 
in Fig.~\ref{fig:freq_3index}.  From top to bottom, the three panels show the result for the single index, 
varying index and multi-component cases respectively.  The smoothly varying and stochastically varying components
are separated, but the 21cm signal is still mixed with the noise.

\subsection{Angular Space Filtering}
\label{sec:angular}
We make angular space filtering to separate the 21cm signal from the noise. 
To simplify the notation, we will omit the frequency variable $\nu$ in the expressions given below, though it 
should be understood that all the observables and beams are functions of $\nu$.

We label the pixels of the sky with the angular position $\vec{\theta}$, then the Gaussian beam 
response $\mat{A}$ is given by
\begin{equation}
A_{\vec{\theta},\vec{\theta}^\prime}= \frac{1}{\sqrt{2\pi}\sigma_\theta} 
e^{-|\vec{\theta}-\vec{\theta}^\prime|^2/(2\sigma_\theta^2)}
\end{equation}

The covariance matrix $\mat{S}$ is given by the angular correlation function of the signal,
\begin{eqnarray}
\mat{S}_{\vec{\theta},\vec{\theta^\prime}}&=&\langle T_{\rm 21}(\vec{\theta}) T_{\rm 21}(\vec{\theta}^\prime) \rangle 
= C(|\vec{\theta}-\vec{\theta}^\prime|)
\label{eq:STT}
\end{eqnarray}
Here we have assumed that the 21cm signal is statistically isotropic and homogeneous, i.e. the 
statistics does not depend on the position in the sky or the direction $\vec{\theta}-\vec{\theta}^\prime$.
Expanding in spherical harmonics,
\begin{equation} 
T(\vec{\theta})=\sum_{l,m} a_{l,m} Y_{l,m}(\vec{\theta})
\label{eq:Tsph}
\end{equation}
Substitute Eq.~(\ref{eq:Tsph}) into Eq.~(\ref{eq:STT}), and using the relation
$\langle a_{l,m} a_{l^\prime,m^\prime}^* \rangle =C_l \delta_{l,l^\prime} \delta_{m,m^\prime}$, we obtain
\begin{equation}
C(|\vec{\theta}-\vec{\theta}^\prime|) = \sum_{l,m} C_l Y_{l,m}(\vec{\theta}) Y_{l,m}^*(\vec{\theta}^\prime) 
\end{equation}
Using the addition theorem for spherical harmonics, 
$$\sum_{m=-l}^l Y_{l,m}(\vec{\theta}) Y_{l,m}^*(\vec{\theta}^\prime) 
= \frac{2l+1}{4\pi} P_l(\vec{\theta} \cdot \vec{\theta}^\prime)$$
where $P_l$ is the Legendre polynomial of $l$-th order,  and $\vec{\theta} \cdot \vec{\theta}^\prime = \cos\theta$ where 
$\theta$ is used to denote the angle between the unit vectors $\vec{\theta}$ and $\vec{\theta}^\prime$,
we finally obtain 
\begin{equation}
\mat{S}_{\vec{\theta},\vec{\theta^\prime}}=\sum_{l} \frac{2l+1}{4\pi} C_l ~P_l(\cos\theta)
\end{equation}

The $C_l$ can be computed from the power spectrum by considering its projection on a thin 
shell with bandwidth $\Delta\nu$ \citep{Santos:2004ju},
\begin{equation}
C_l(\nu)=\frac{2 \Delta \nu^2}{\pi}\int k^2 d k P_{\rm 21}(k,\nu) j_l^2[k r(\nu)]
\end{equation}

\begin{figure}[htbp]
\centering
\includegraphics[width=0.45\textwidth]{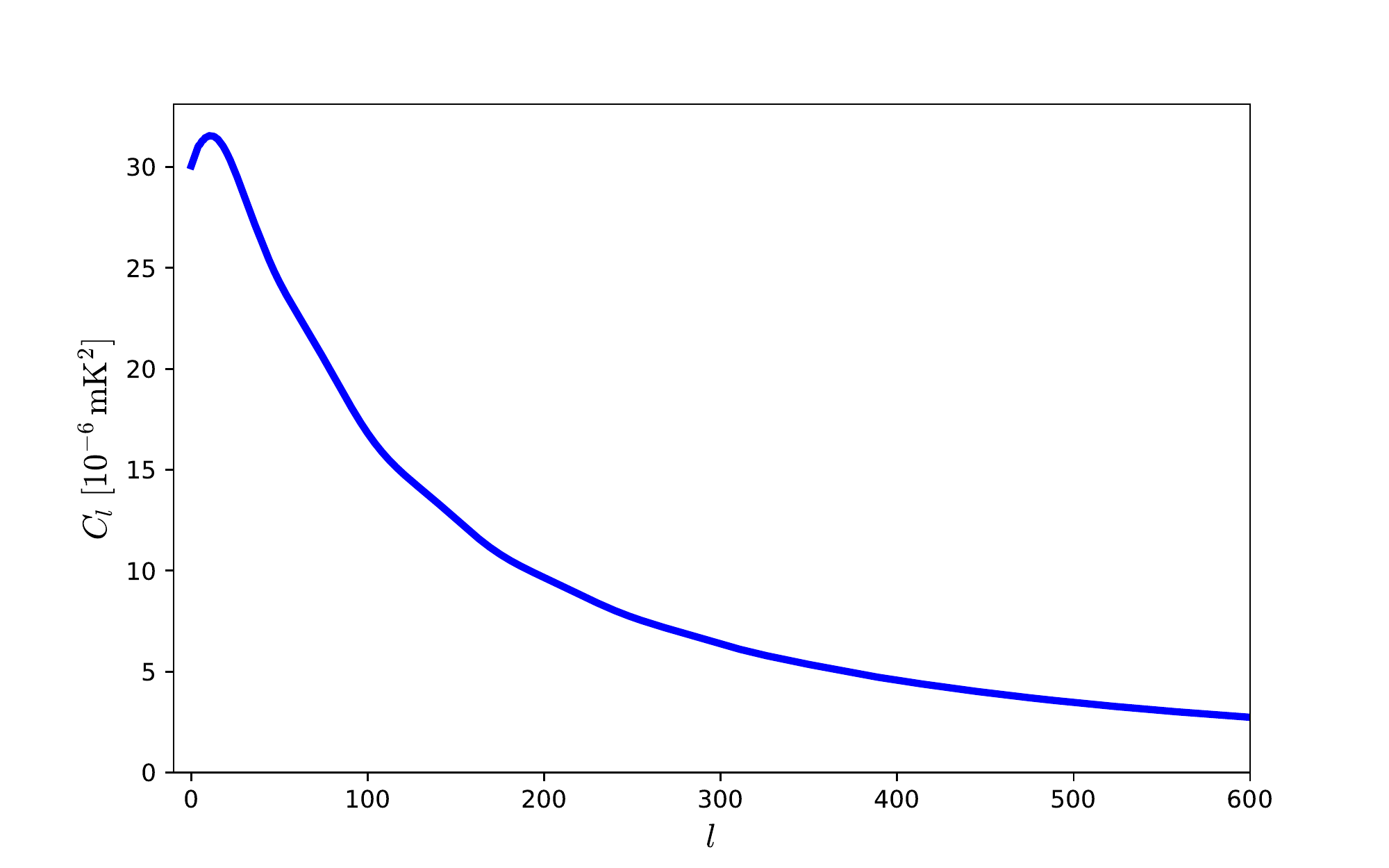}
\includegraphics[width=0.45\textwidth]{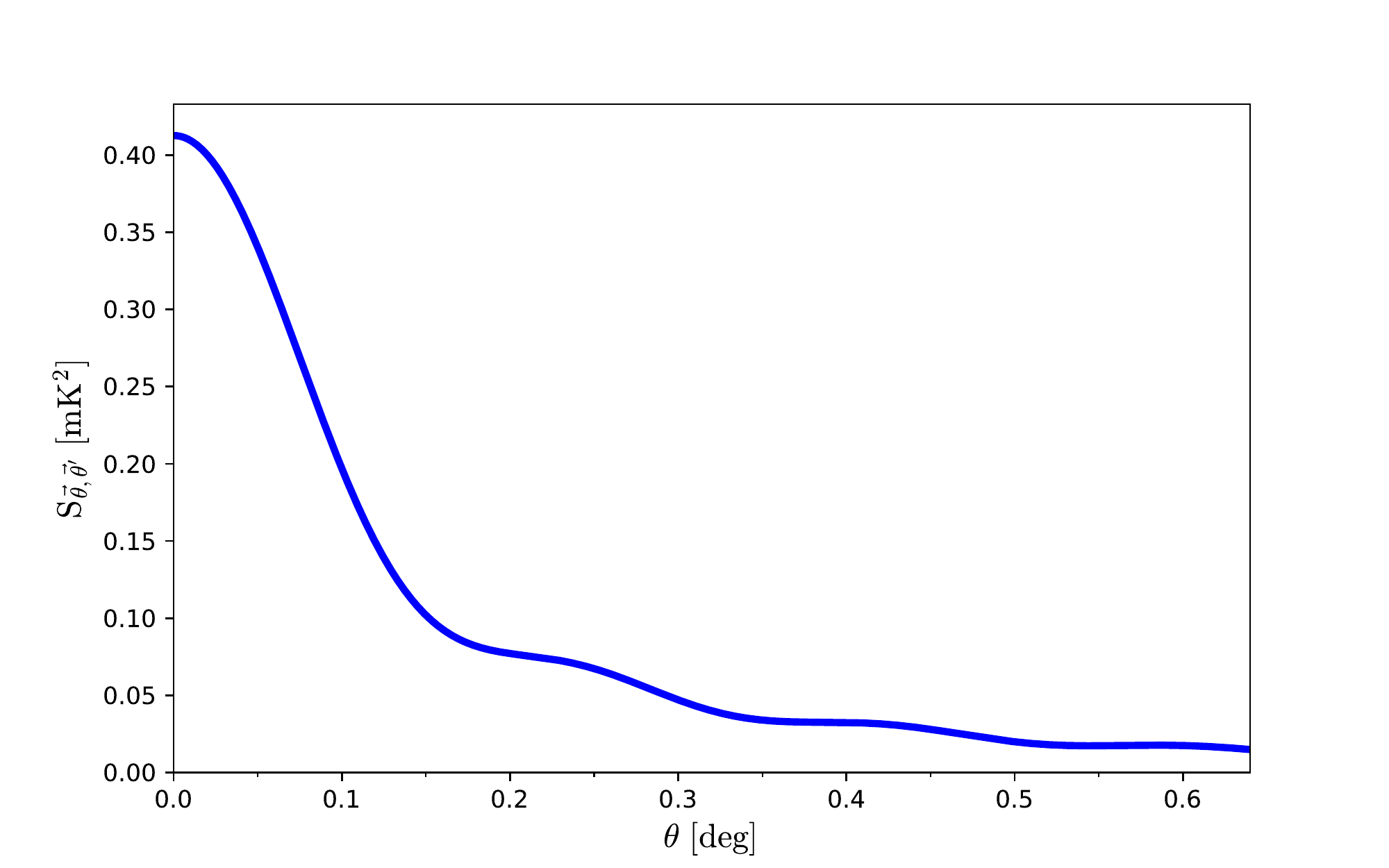}
\caption{ $C_l$ (top) and $C(\theta)$ of 21cm signal at $z\sim 0.8$. }
\label{fig:Cl}
\end{figure}
In Fig.~\ref{fig:Cl} we show the angular power spectrum $C_l$ (top panel) and corresponding angular correlation 
function $C(\theta)$ (bottom) panel. 
The correlation between 21cm signal drops rapidly at  $l \sim 10^2$ or degree scale.

The noise covariance matrix for pixels $\vec{\theta},\vec{\theta}^\prime$  is given by 
\begin{equation}
\mat{N}_{\vec{\theta},\vec{\theta}^\prime}= N_0 \delta_{\vec{\theta},\vec{\theta}^\prime}, 
\end{equation}
where for simplicity we have assumed a constant noise $N_0$.  Note that in Eq.~(\ref{eq:y}),
if the vector $\mat{x}$ denoted sky pixels while $\mat{y}$ denotes time-ordered data, 
then the different elements of $\mat{n}$ are data taken at different time and may be considered independent random
samples, so the noise matrix $\mat{N}$ is diagonal. If we use pixels finer than the beam size, then
when we re-bin the time-ordered data into sky pixels, the noise in adjacent pixels with angular distance smaller than the 
beam size would be correlated, this is automatically taken into account  in the Wiener filter Eq.~(\ref{eq:wiener}) by the 
response matrix $\mat{A}$ and $\mat{A}^T$. In the real world, the noise may be more complicated,  for example, 
the noise level may be direction-dependent, either due to  brighter sky temperature, or because the operating condition of
the telescope receiver. Furthermore, even the data taken at different time may have some correlation due to the presence
of $1/f$ noise. These effects can also be handled by the Wiener filtering algorithm, if the noise covariance matrix
$\mat{N}$ is known. In the present work we shall assume the simple case where the noise is uncorrelated and constant.

\begin{figure}[htbp]
\centering
\includegraphics[width=0.46\textwidth]{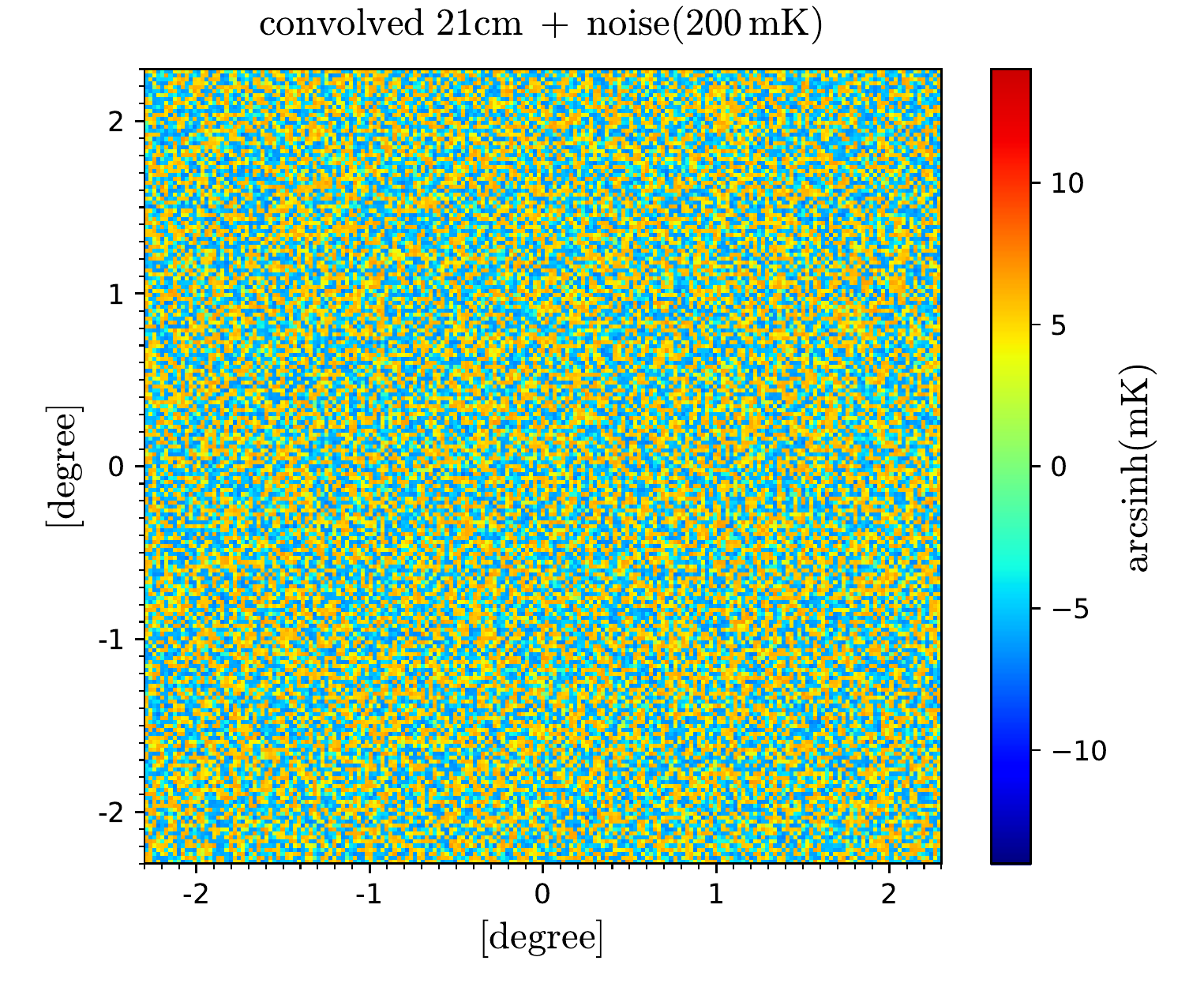}
\includegraphics[width=0.46\textwidth]{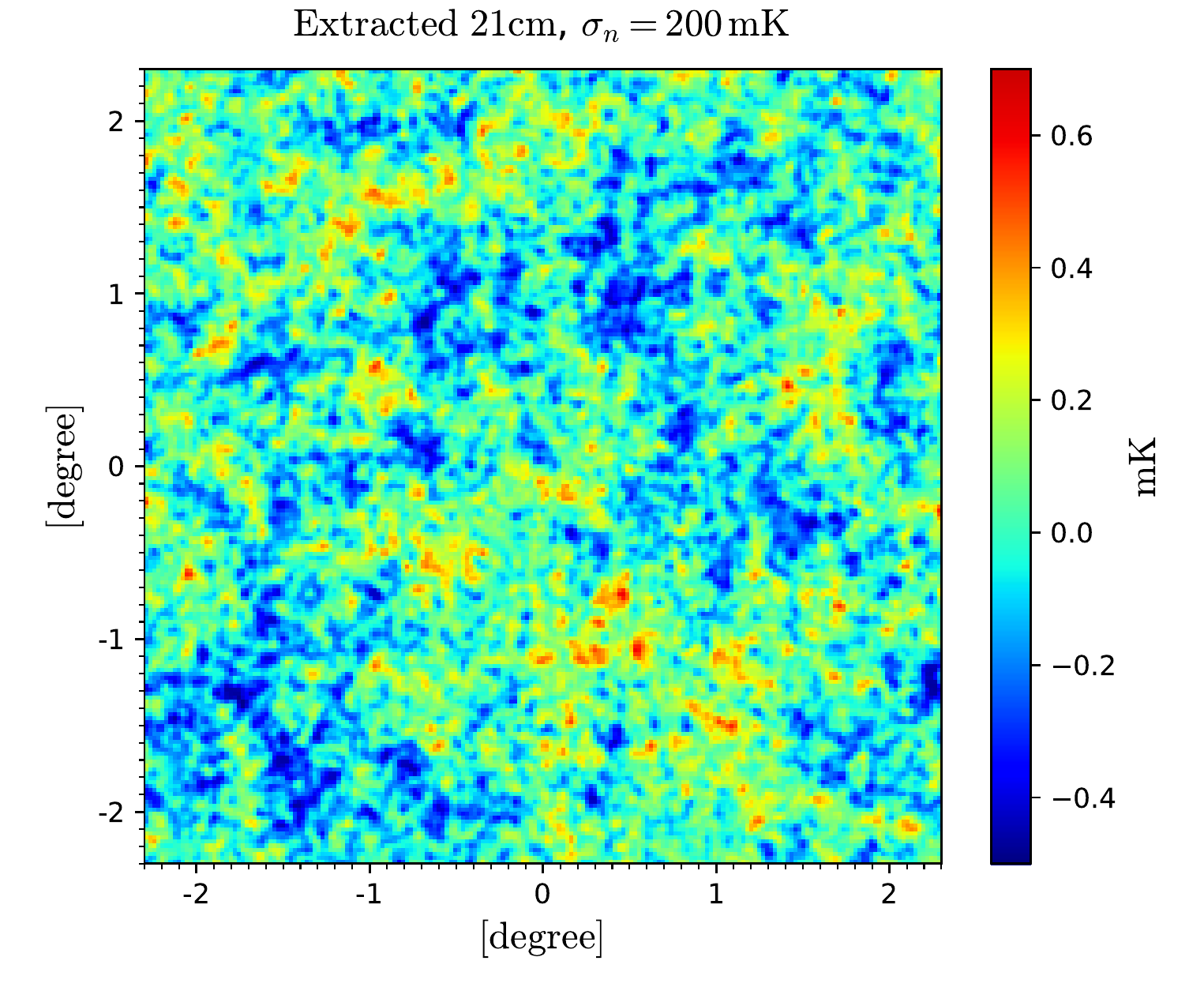}
\includegraphics[width=0.46\textwidth]{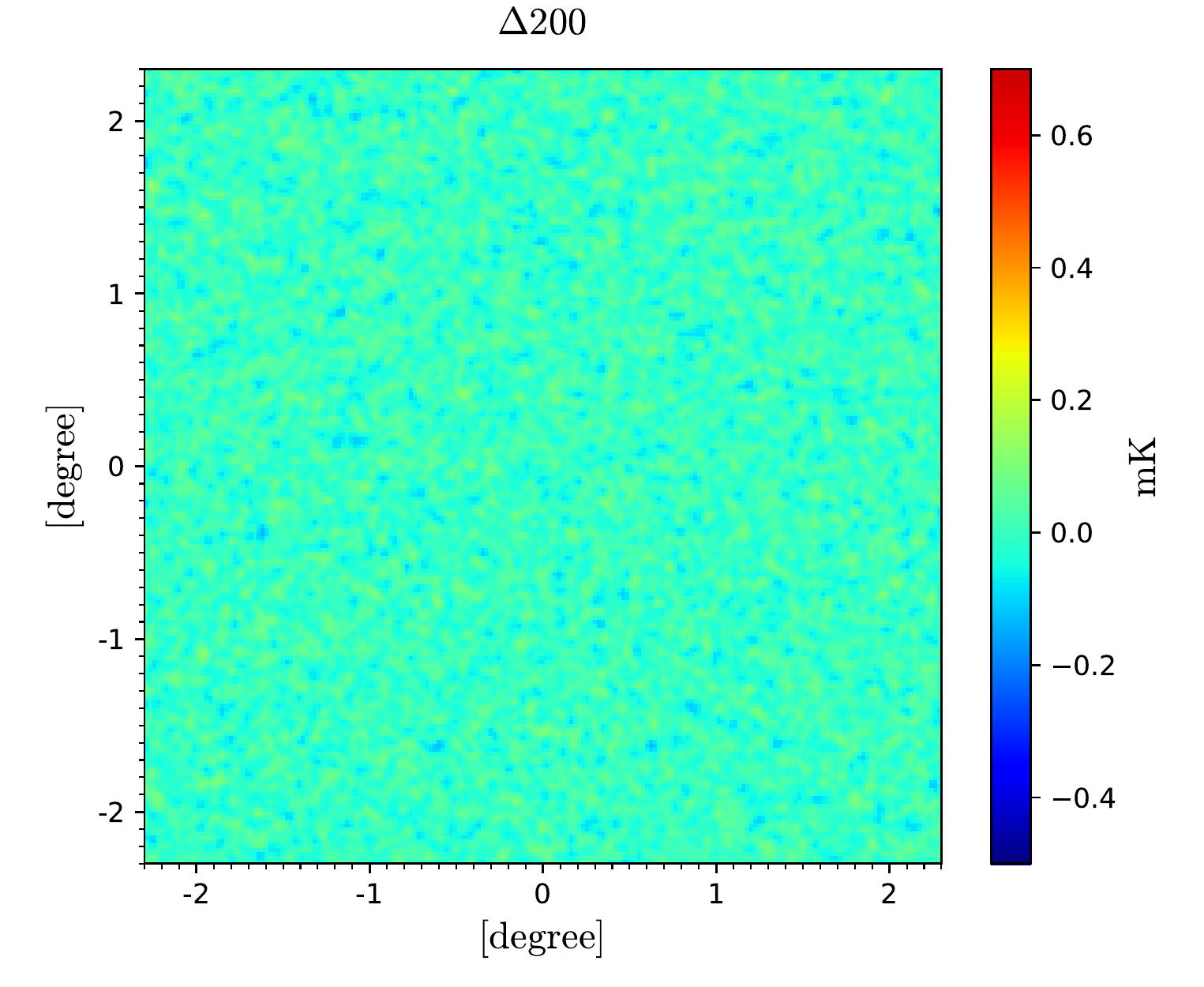}
\caption{ Top Panel: 21cm map + noise; Middle Panel: extracted 21cm map;
Bottom: difference between the extracted and input 21cm maps. }
\label{fig:pix_ext21}
\end{figure}
\begin{figure}[htbp]
\centering
\includegraphics[width=0.46\textwidth]{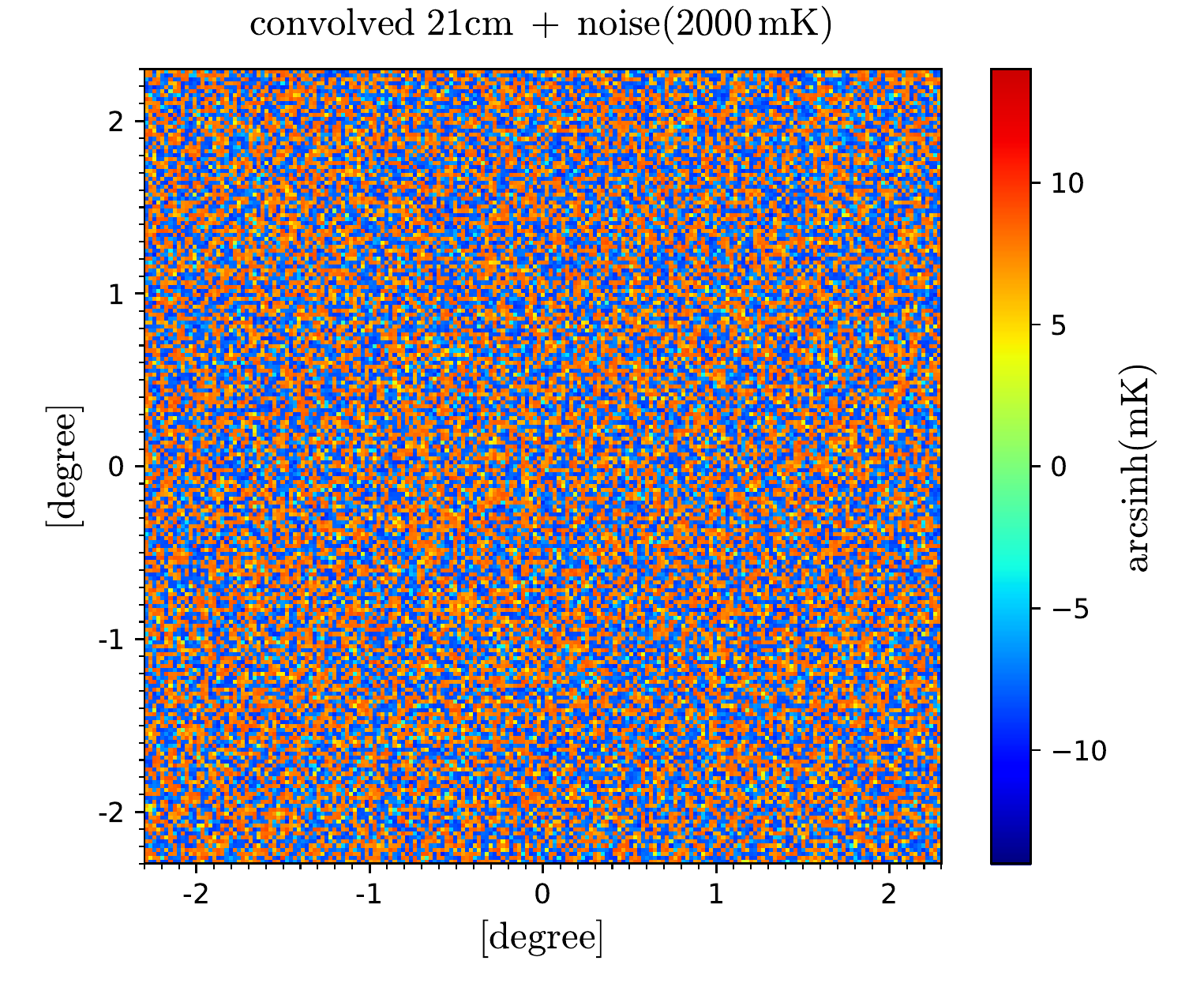}
\includegraphics[width=0.46\textwidth]{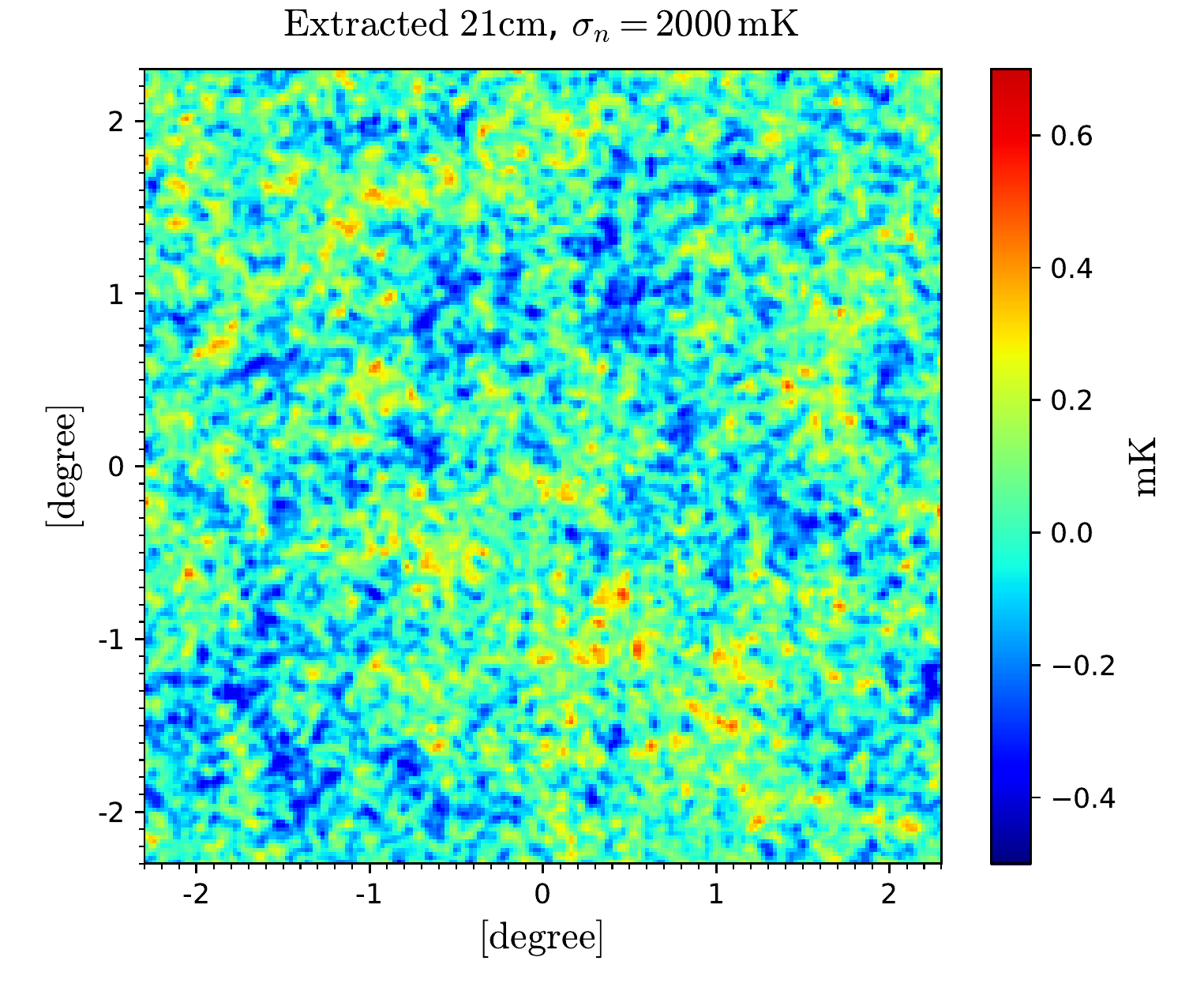}
\includegraphics[width=0.46\textwidth]{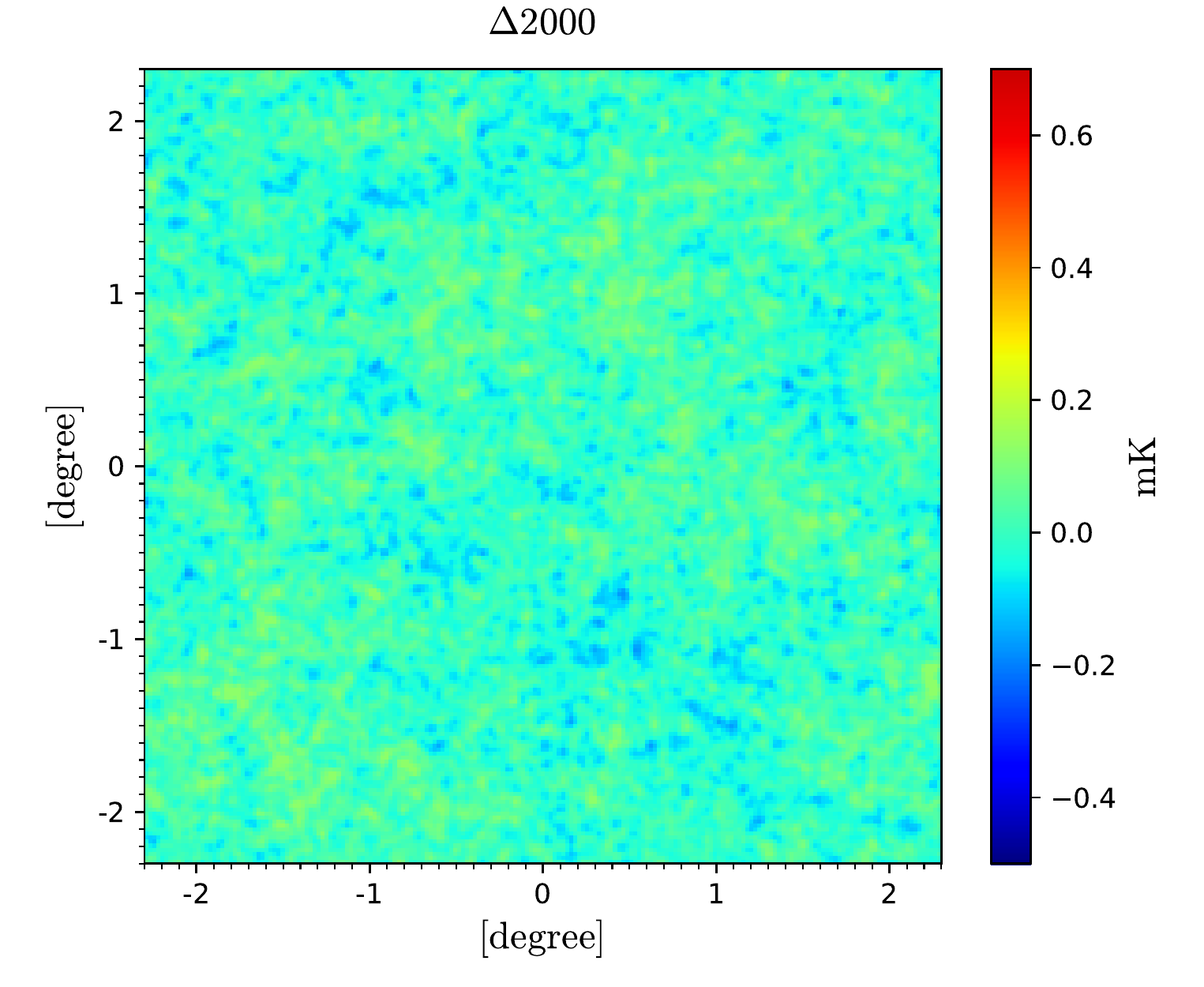}
\caption{ Same as Fig.~\ref{fig:pix_ext21}, but with higher noise level (2000 mK).  }
\label{fig:map_highnoise}
\end{figure}

Fig.~\ref{fig:pix_ext21} shows the extracted 21 cm map by applying the Wiener filter. From top to bottom,  we plot
the input map with 21cm signal plus 200 mK noise,  the extracted 21cm map, and 
the difference between the extracted map and the input map. Despite of the high noise level, the 21cm signal is successfully
recovered, the difference between the recovered map and the original one is very small. 
Note that the Wiener filter is obtained by using the angular correlation function computed from the cosmological model, 
which corresponds to the ensemble average value, the actual realization may differ slightly due to sample variance, so the 
recovery would not be perfect.  

In fact, the method could also work reasonably well even if the noise level is still higher. This is shown
in Fig.~\ref{fig:map_highnoise}, where the noise is assumed to be 2000 mK. Here we see that the difference between the 
recovered 21cm map and the original one is larger than in Fig.~\ref{fig:pix_ext21}, but the overall structure of the 21cm 
intensity is still clearly seen, and the difference between the two maps is still much smaller than the 21cm brightness
temperature.

\begin{figure}
\centering
\includegraphics[width=0.46\textwidth]{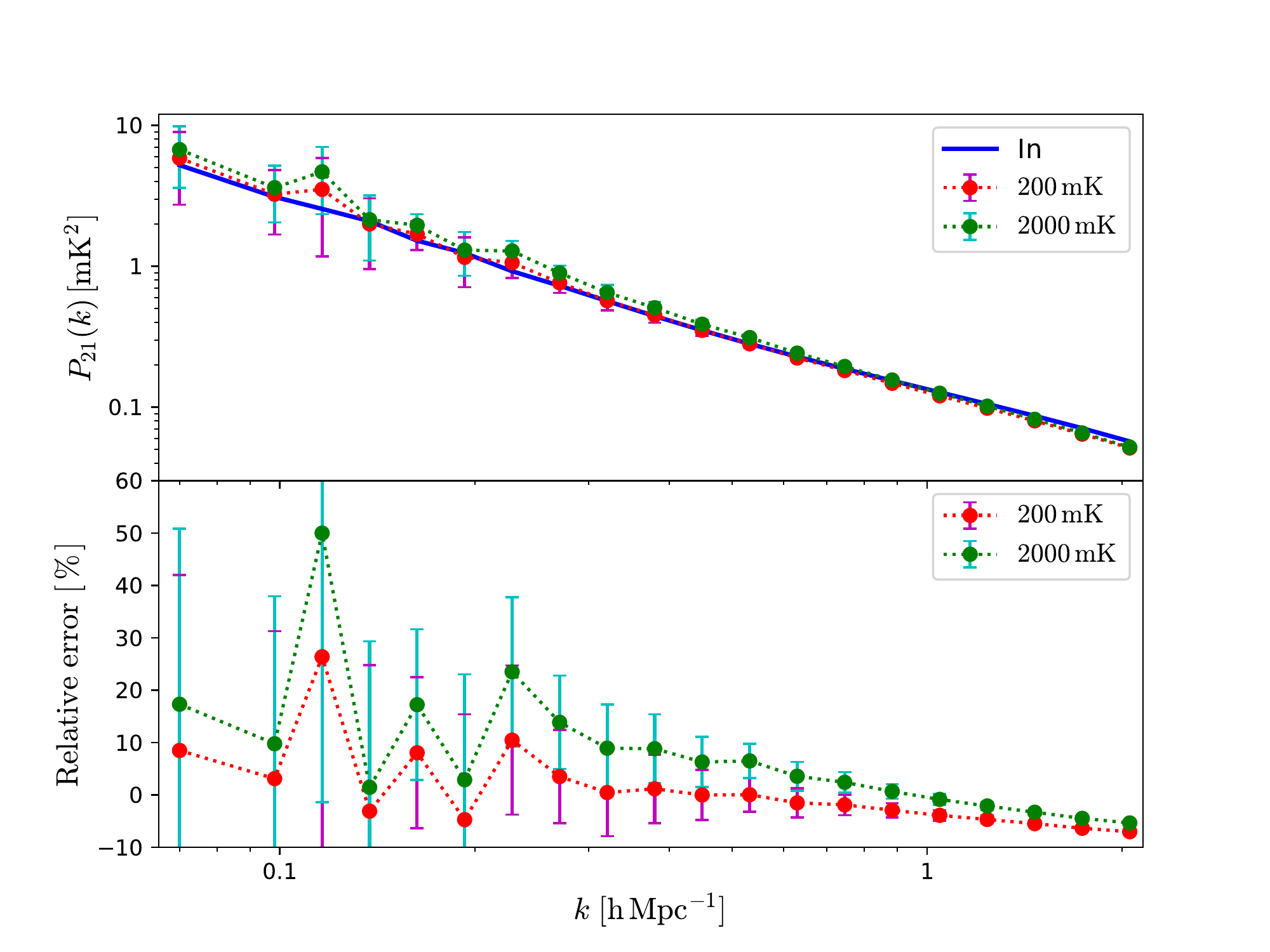}
\caption{The recovered 21cm power spectrum (top panel) and the relative error (bottom panel) for this simulation box. 
We show the power for the original (blue line) signal, the reconstruction with 200 mK noise (red symbols), and 2000 mK 
noise (green symbols) in the figure.}
\label{fig:powerspec}
\end{figure}

In Fig.~\ref{fig:powerspec} we show the recovered 21cm power spectrum and the relative error. The error bars are 
estimated from the variance in $k$-space. The error obtained here is for the simulation box which is  $(86 \Mpc ~h^{-1})^3$.
If we assume that the relative error scales simply as $\Delta P/P \sim V^{-1/2}$, 
we estimate that in order to achieve 1\% statistical precision on power spectrum at $k=0.1 \Mpc ~h^{-1}$, the 
required survey volume is $(370 \Mpc ~ h^{-1})^3$. The actual error may be larger when sampling variance and 
imperfection in reconstruction are taken into account.

\begin{figure*}
\centering
\includegraphics[width=0.46\textwidth]{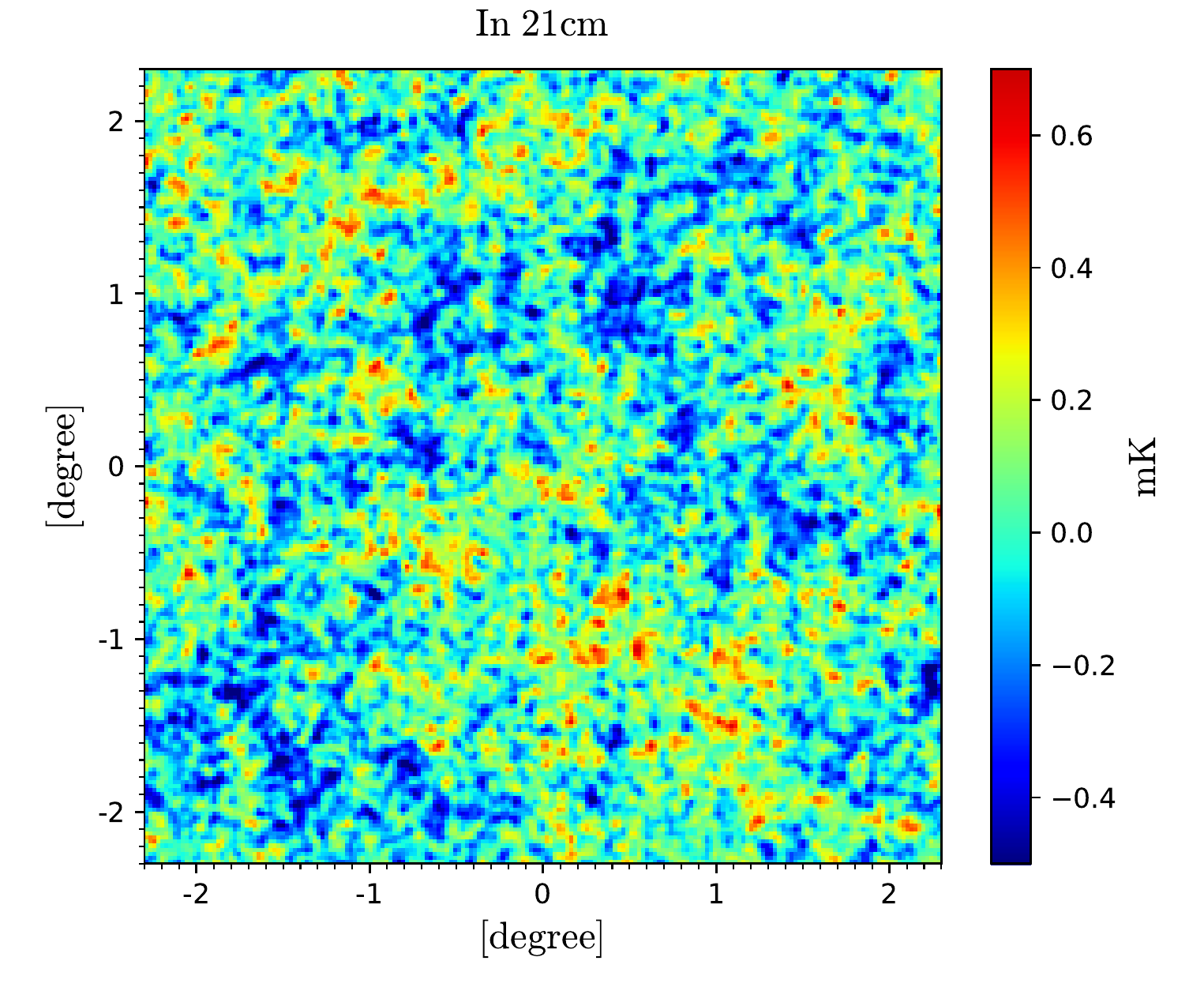}
\includegraphics[width=0.46\textwidth]{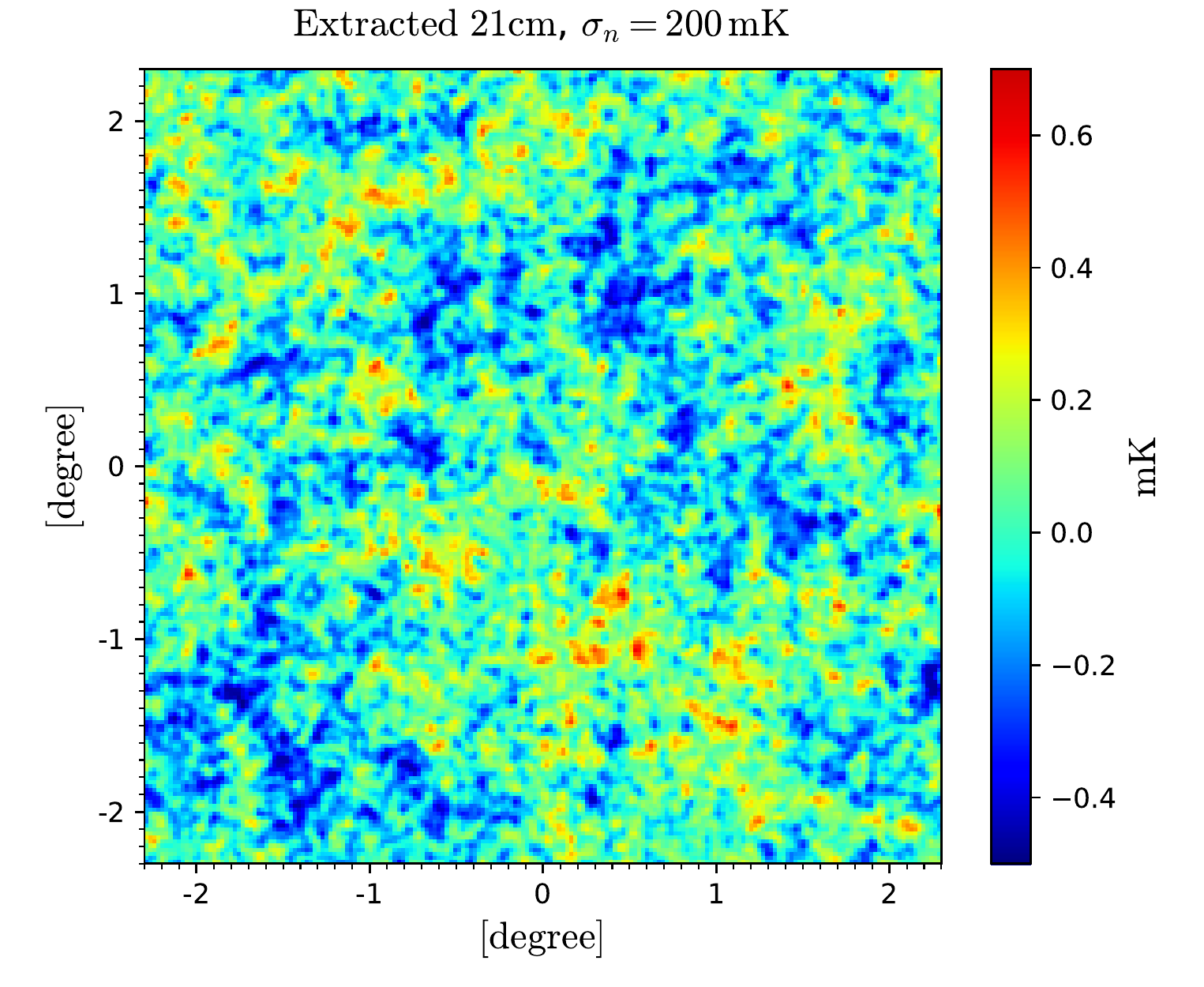}\\
\includegraphics[width=0.46\textwidth]{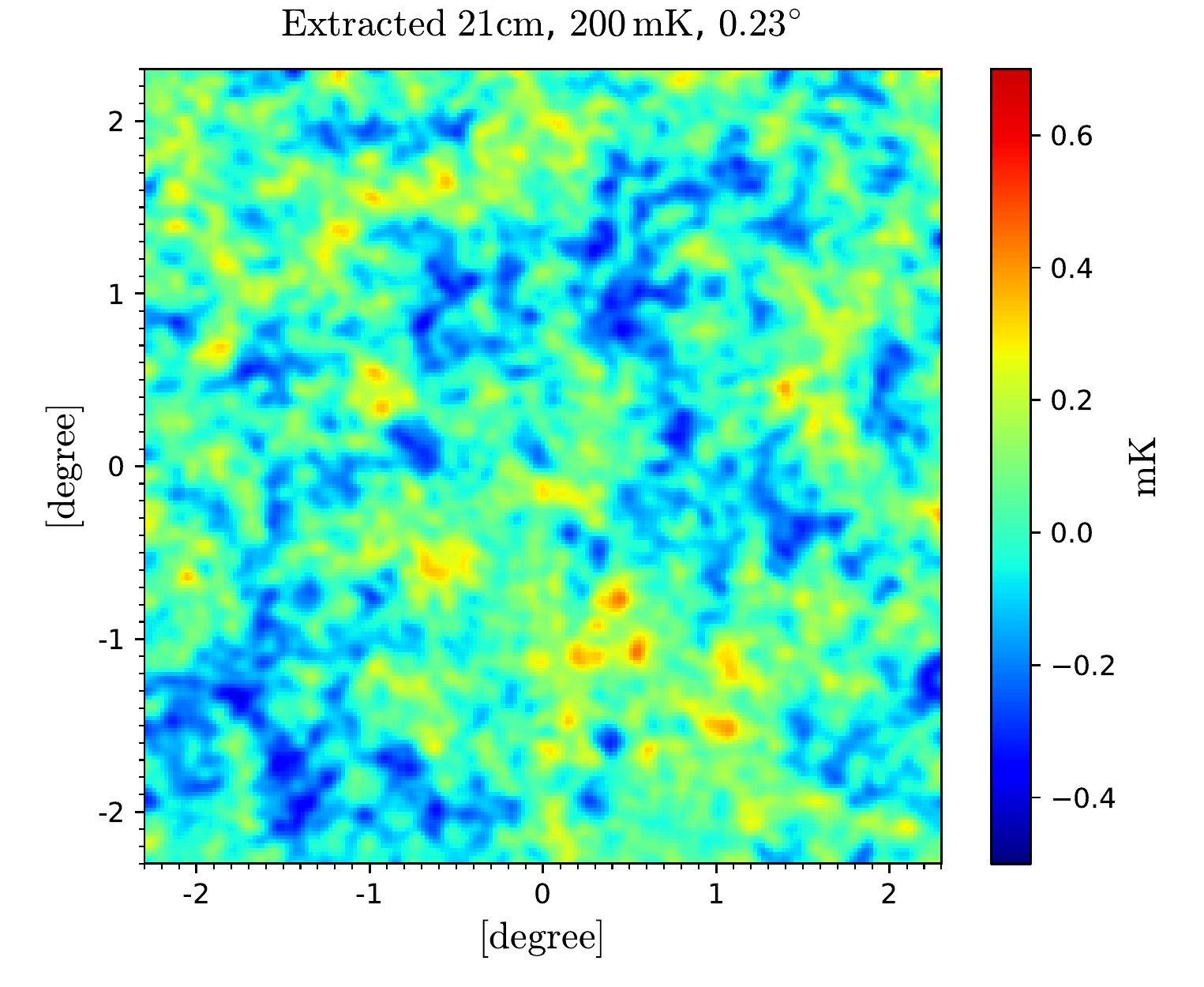}
\includegraphics[width=0.46\textwidth]{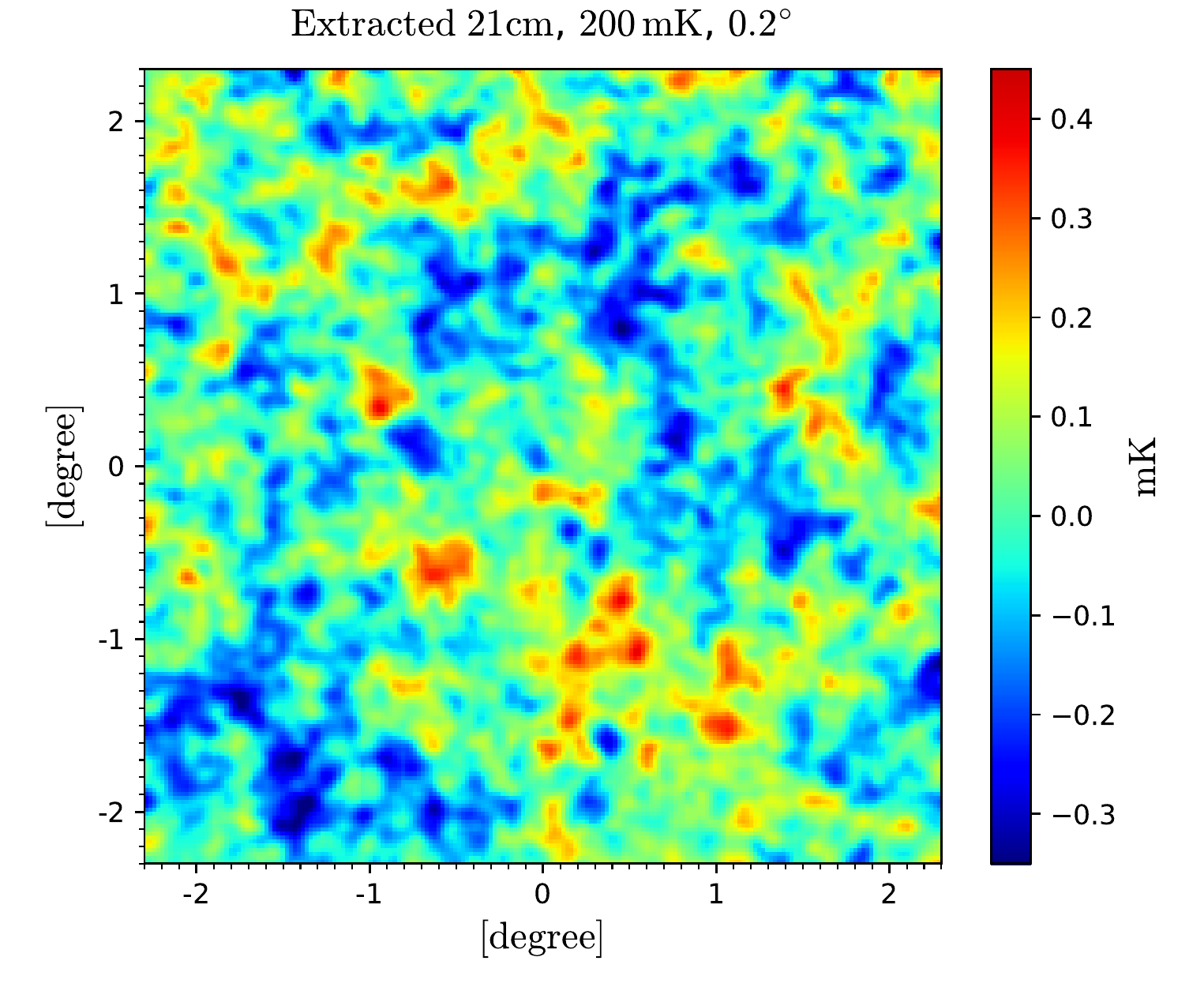}
\caption{Effect of inaccurate beam on map reconstruction. We assume here the real beam size is $0.25^\circ$. 
Top Left: input 21cm map; Top Right: reconstructed 21cm map from noise of 200 mK, 
with the correct beam size of $0.25^\circ$; Bottom Left: reconstruction with the incorrect beam width of $0.23^\circ$;
Bottom Right: reconstruction with the incorrect beam width of $0.20^\circ$.}
\label{fig:pix_ext21}
\end{figure*}

When making the map, the Wiener filter automatically deconvolves the data. 
In the above analysis the beam is assumed to be perfectly known, so the recovery is very accurate even when noise is 
present. Indeed, some finer details whose scales are smaller than that of the beam width are recovered in the 
reconstructed map, which shows the power of the Wiener filter method.
However, in the real world the beam is only known either by electromagnetic field simulation, 
or by calibration measurements, and both approaches have errors. 
We make a simple demonstration of the effect of inaccurately known beam by the following excise: 
we assume the real beam is an Gaussian with a beam width (FWHM) of $0.25^\circ$, 
but then in the reconstruction we use Gaussian beams with slightly different beam widths. The result is shown 
in Fig.~\ref{fig:pix_ext21}. The top panels show the original 21cm signal (top left, same as the top panel of 
Fig.~\ref{fig:inmap}) and the reconstructed map with the correct beam size (top right, same as the middle panel of 
Fig.~\ref{fig:pix_ext21}), we put these here again for easy comparison. 
The bottom panels show the reconstructed map with Gaussian beams of incorrect 
beam widths of $0.23^\circ$ (Bottom Left) and $0.20^\circ$ (Bottom Right). We see that when the incorrect beam 
widths are used, the whole reconstructed map become "fuzzier", the finer details of the original maps are lost, 
though the overall large scale structure are still very similar. In the real world, of course, the deviation from the beam 
might be more irregular and complicated, but the overall effect would be losing the details below the beam resolution while
retaining the larger overall structures.


\section{Discussions and Conclusion}
\label{sec:conclusion}

A 3D cosmological neutral hydrogen survey over a large fraction of the sky is an efficient way to study our universe.  
A number of instruments have been developed or are being designed for such surveys. However, a great challenge is 
that both the foreground radiation and the noise are several orders of magnitude larger than the 21cm signal.
To extract the cosmological 21cm signal from the data collected from such instruments, 
an efficient extraction method is required. 

This paper is an exploratory study of this issue using the Wiener filter, 
which is widely utilized in signal processing field.  
We have taken as an example the analysis of data processing for a mid-redshift experiment, which is aimed at 
measuring the dark energy equation of state by using the BAO features in the large scale structure. However, the method
is also applicable for the EoR experiments.
Assuming that the data has been pre-processed, 
and an image cube have been produced,  we used a two step procedure to extract the 21cm 
signal. We first subtract the foreground by removing the smooth component in the frequency 
spectrum along each line of sight. 
Previously, \citet{2012MNRAS.419.3491L} applied the Wiener filter method to extract the foreground from 
21cm experiment, but they are mostly concerned mostly with the frequency spectrum, which is applicable to 21cm global 
spectrum experiment, or to one line of sight for the 21cm intensity mapping experiment, corresponding to this first step.
However, we then go one step further, extracting the 21cm signal by applying the Wiener filter on the two dimensional 
angular space.
Our simulation show that the 21cm signal could be recovered with good precision.  
In actual data analysis, the power spectrum of the 21cm signal is not precisely known, but from other cosmological 
observations approximate value could be inferred. Starting from an approximate value, one can apply the filter iteratively 
to improve the estimate.

In the present study we have made a number of simplifying assumptions. In an actual experiment, 
the beam shape is more complicated, frequency-dependent and only known to a limited precision, 
the calibration procedure may introduce additional errors, and the noise may be non-thermal and have more
complicated statistical properties. All of these factors may affect the extraction of the 21cm signal. To overcome 
these problems, one needs to consider the specific experiment. Nevertheless, the Wiener filtering may provide 
a very useful tool for 21cm data analysis.


\acknowledgements
This research is supported by the Ministry of Science and Technology grant 2016YFE0100300, NSFC grants 11473044, U1501501,
U1631118, 11633004, and CAS grant QYZDJ-SSW-SLH017. F. Q. Wu also acknowledge the support by the CSC 
Cai Yuanpei grant.

\bibliographystyle{aasjournal}
\bibliography{wiener}
\end{document}